\documentclass[review,number,sort&compress,longtitle]{elsarticle}
\usepackage{graphicx,amsmath,amsfonts,amscd,amsthm,amssymb,pstricks}%,lineno}
\makeatletter
\newcommand*{\Hy@backout}[1]{}
\makeatother

\newcommand{\dif}[1]{\ensuremath{\mathrm{d}#1}}
\newcommand{\Int}[3]{\ensuremath{\int\limits_{#1}^{#2}\!\dif{#3}\;}}

\newcommand{\Exp}[1]{\ensuremath{\mathrm{e}^{#1}}}

\begin{document}
\begin{frontmatter}
\title{Description of the luminosity evolution for the CERN LHC including dynamic aperture effects. Part II: application to Run~1 data\tnoteref{mytitlenote}}
\tnotetext[mytitlenote]{Research supported by the HL-LHC project}
\author[cern]{M. Giovannozzi\corref{mycorrespondingauthor}}
\cortext[mycorrespondingauthor]{Corresponding author}
\ead{massimo.giovannozzi@cern.ch}
\author[cern]{F.F.~Van~der~Veken}
\address[cern]{Beams Department, CERN, CH 1211 Geneva 23, Switzerland}
\begin{abstract}
In recent years, modelling the evolution of beam losses in circular proton machines starting from the concept of dynamic aperture its time evolution has been the focus of intense research. Results from single-particle non-linear beam dynamics have been used to build simple models that proved to be in good agreement with beam measurements. These results have been generalised, thus opening the possibility to describe also the luminosity evolution in a circular hadron collider. In a companion paper~\cite{lumi_Part_I}, the derivation of a scaling law for luminosity, which includes both burn off and pseudo-diffusive effects, has been carried out. In this paper, the proposed models are applied to the analysis of the data collected during the CERN Large Hadron Collider (LHC) Run~1. A data set referring to the proton physics runs for the years 2011 and 2012 has been analysed and the results are proposed and discussed in detail in this paper.
\end{abstract}
\begin{keyword}
Dynamic aperture \sep luminosity evolution \sep LHC 
\end{keyword}
\end{frontmatter}
%
%\linenumbers
% 
\section{Introduction}
The unavoidable non-linear magnetic field errors that plague the dynamics of charged particles of modern superconducting colliders, inducing new and potential harmful effects, require the development of new approaches to perform more powerful analyses and to gain insight in the beam dynamics. It is the case of the critical revision of the concept of dynamic aperture (DA) and of its dependence on time~\cite{dynap1,dynap2} for the case of single-particle effects. Such a scaling law was later successfully extended to the case in which weak-strong beam-beam effects are taken into account~\cite{loginvb-b}. More importantly, this scaling law paved the way to describe the time evolution of beam losses in a circular particle accelerator under the influence of non-linear effects~\cite{lossesPRSTAB}, verified experimentally using data from CERN accelerators and the Tevatron, which is at the heart of a novel method to measure experimentally the DA in a circular ring~\cite{DAexp_nekor}.

The description of the luminosity evolution in a circular collider profited from this novel framework. The first attempts to derive a new model are reported in~\cite{Lumi_fit,IPAC14}, whereas a more complete and accurate modelling is described in~\cite{lumi_Part_I}. The proposed approach is put in action by analysing a selection of luminosity data collected at the LHC during Run~1, in particular for the proton physics runs in the years 2011 and 2012, which is the focus on this paper, where the result of these analyses is discussed in detail. Furthermore, it is worth mentioning that the scaling law~\cite{lossesPRSTAB} has been used also in the analysis of dedicated beam-beam experiments performed at the LHC~\cite{MC1,MC2}.

Two points are worth stressing. Firstly, the focus of this paper is a test of the descriptiveness of the novel model, which is probed by checking the quality of the agreement with the LHC data. Note that the issue of predictiveness will be addressed in a different paper. Secondly, the arguments used to build the proposed model are rather general, thus implying that they should not be applicable to LHC only, but to circular colliders in general.

It is recalled that the starting point is the expression of luminosity, which is a key figure-of-merit for colliders that, neglecting the hourglass effect, reads~\cite{Herr}
\begin{equation}
L = \frac{\gamma_{\rm r} \, f_{\rm rev} \, k_{\rm b} \, n_1 \, n_2}%
{4 \, \pi \epsilon^* \beta^*} \, F(\theta_{\rm c}, \sigma_{z}, \sigma^* ),
\label{lumi}
\end{equation}
where $\gamma_{\rm r}$ is the relativistic $\gamma$-factor, $f_{\rm rev}$ the revolution frequency, $k_{\rm b}$ the number of colliding bunches, $n_{\rm i}$ the number of particles per bunch in each colliding beam, $\epsilon^*$ is the RMS normalised transverse emittance, and $\beta^*$ is the value of the beta-function at the collision point. The total beam population is defined as $N_{j}=k_{\rm b} \, n_{j}$. Different bunches have different collision schedules, meaning that they collide in different interaction points. Hence, the following analysis could have been performed on a bunch-by-bunch basis and considering bunches with the same collision schedule as members of the same class. However, a simplified approach has been applied, namely the total intensity $N_{j}$ has been rescaled by
$\frac{k_{\rm b, ATLAS, CMS}}{k_b}\, ,$
where $k_{\rm b, ATLAS, CMS}$ represents the number of bunches colliding in the two high-luminosity experiments. The underlining assumption is that the effects generated by the collisions in the two other low-luminosity experiments can be neglected. This is supported by the difference in typical peak luminosities, whose ratio to those of the high-luminosity experiments is $\sim 10^{-1}, \sim 10^{-3}$ for LHCb and Alice in 2011, respectively and $\sim 10^{-2}, \geq 10^{-3}$ for LHCb and Alice in 2012, respectively~\cite{lpc}.

The factor $F$ accounts for the reduction in volume overlap between the colliding bunches due to the presence of a crossing angle and is a function of  half the crossing angle $\theta_{\rm c}$ and the transverse and longitudinal RMS dimensions $\sigma^*, \sigma_{z}$, respectively according to~\cite{Herr}:
\begin{equation}
F(\theta_{\rm c}, \sigma_{z}, \sigma^* )=\frac{1}{\sqrt{1+
\left ( \displaystyle{\frac{\theta_{\rm c}}{2} \, \frac{\sigma_{z}}{\sigma^*}} 
\right )^2}} \, .
\label{geofac}
\end{equation}
Note that $\sigma^*=\sqrt{\beta^* \, \epsilon^*/(\beta_{\rm r} \, \gamma_{\rm r})}$, where $\beta_{\rm r}$ is the relativistic $\beta$-factor. Equation~(\ref{lumi}) is valid in the case of round beams ($\epsilon_x^*=\epsilon_y^*=\epsilon^*$) and round optics ($\beta_x^* = \beta_y^*=\beta^*$). For our scope, Eq.~(\ref{lumi}) will be recast in the following form~\cite{Herr}:
\begin{equation}\label{LumiDef}
L = \Xi \, N_1 \, N_2,  \qquad  \Xi = \frac{\gamma_{\rm r} f_{\rm rev}}%
{4 \, \pi \epsilon^* \beta^* \, k_{\rm b} } 
F(\theta_{\rm c}, \sigma_{z}, \sigma^* )
\end{equation}
in which the dependence on the total intensity of the colliding beams is highlighted and the other quantities are included in the term $\Xi$ .

The plan of the paper is the following: in section~\ref{sec:data} a global overview of the LHC data is presented, including a discussion of the time-dependence of some beam parameters, and the choice made for the data analysis. The application of the proposed model is presented in three different sections each dealing with one of the three main observables considered, namely the time evolution of the luminosity over a physics fill (section~\ref{sec:lumi_data}), the integrated luminosity over a physics fill (section~\ref{sec:int_lumi_data}), and the optimal duration of a physics fill (section~\ref{sec:fill_duration_data}). Conclusions are drawn in section~\ref{sec:conclusions}, whereas the detailed discussion of the important features of the numerical models proposed in this paper is presented in \ref{app:difficulties}.
\section{LHC data from Run~1}\label{sec:data}
\subsection{General considerations}\label{sec:generalities}
\begin{figure}[p]
  \begin{center}
    \begin{tabular}{@{}c@{}@{}c@{}}
      \includegraphics[width=0.49\linewidth,clip=]{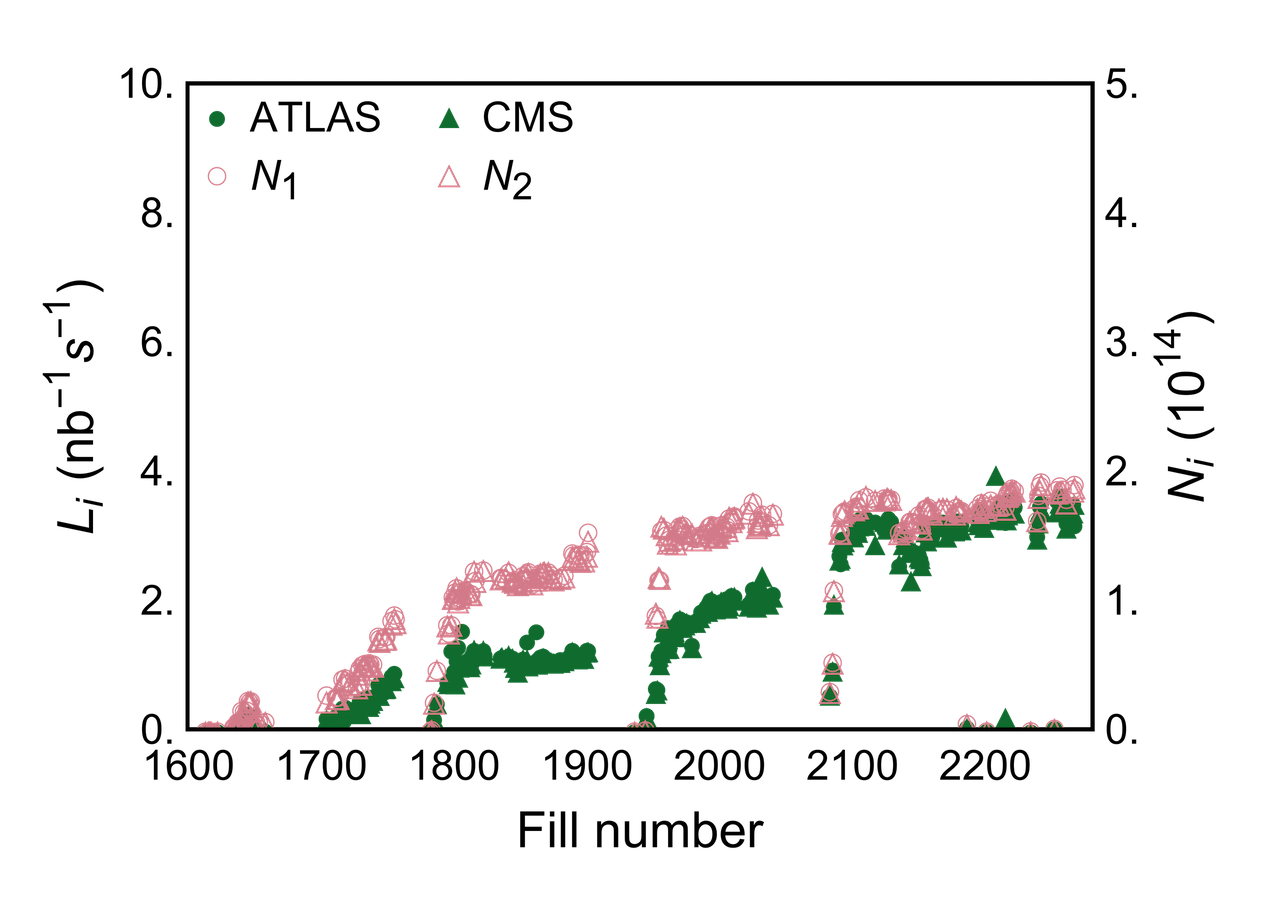} &
      \includegraphics[width=0.49\linewidth,clip=]{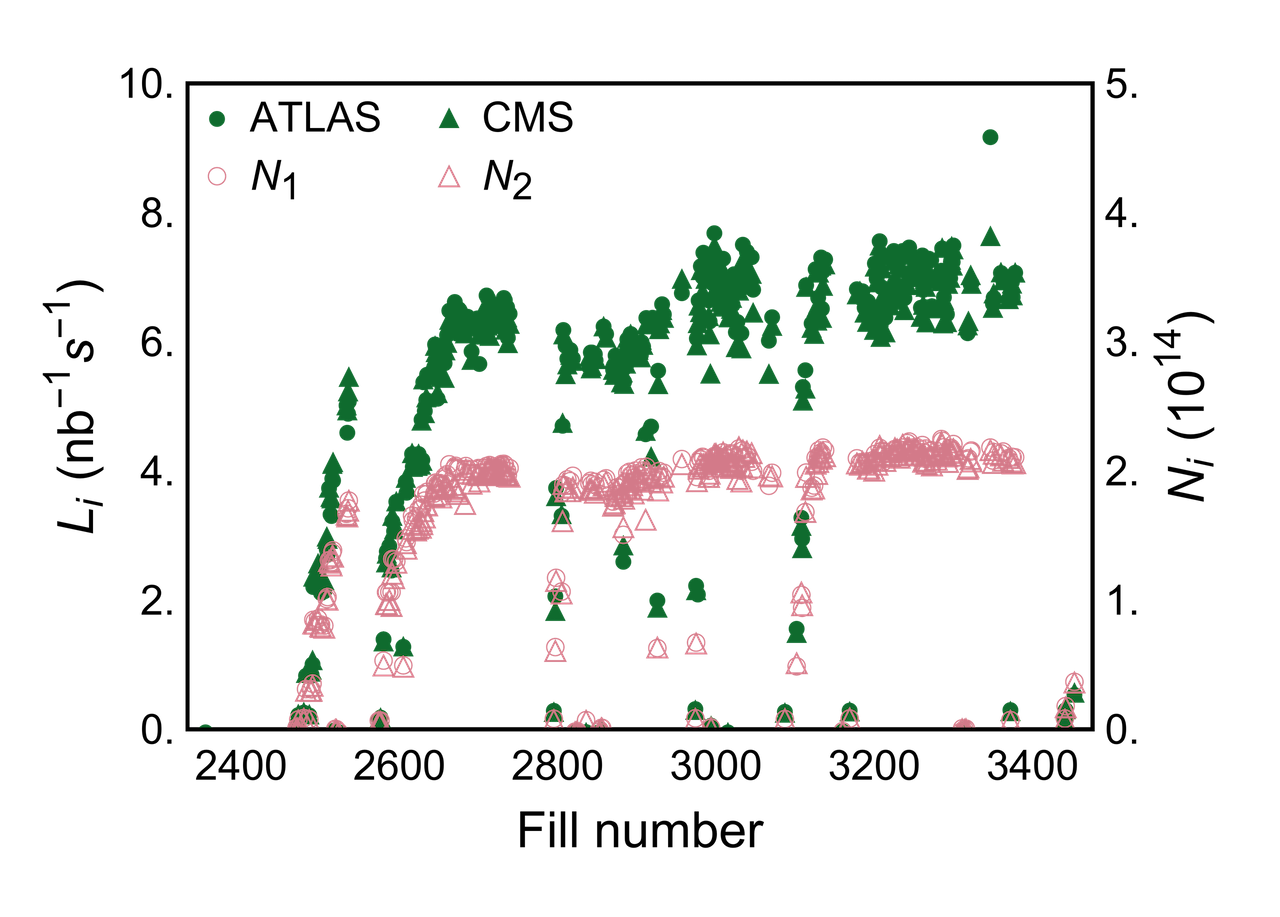} \\
      \includegraphics[width=0.49\linewidth,clip=]{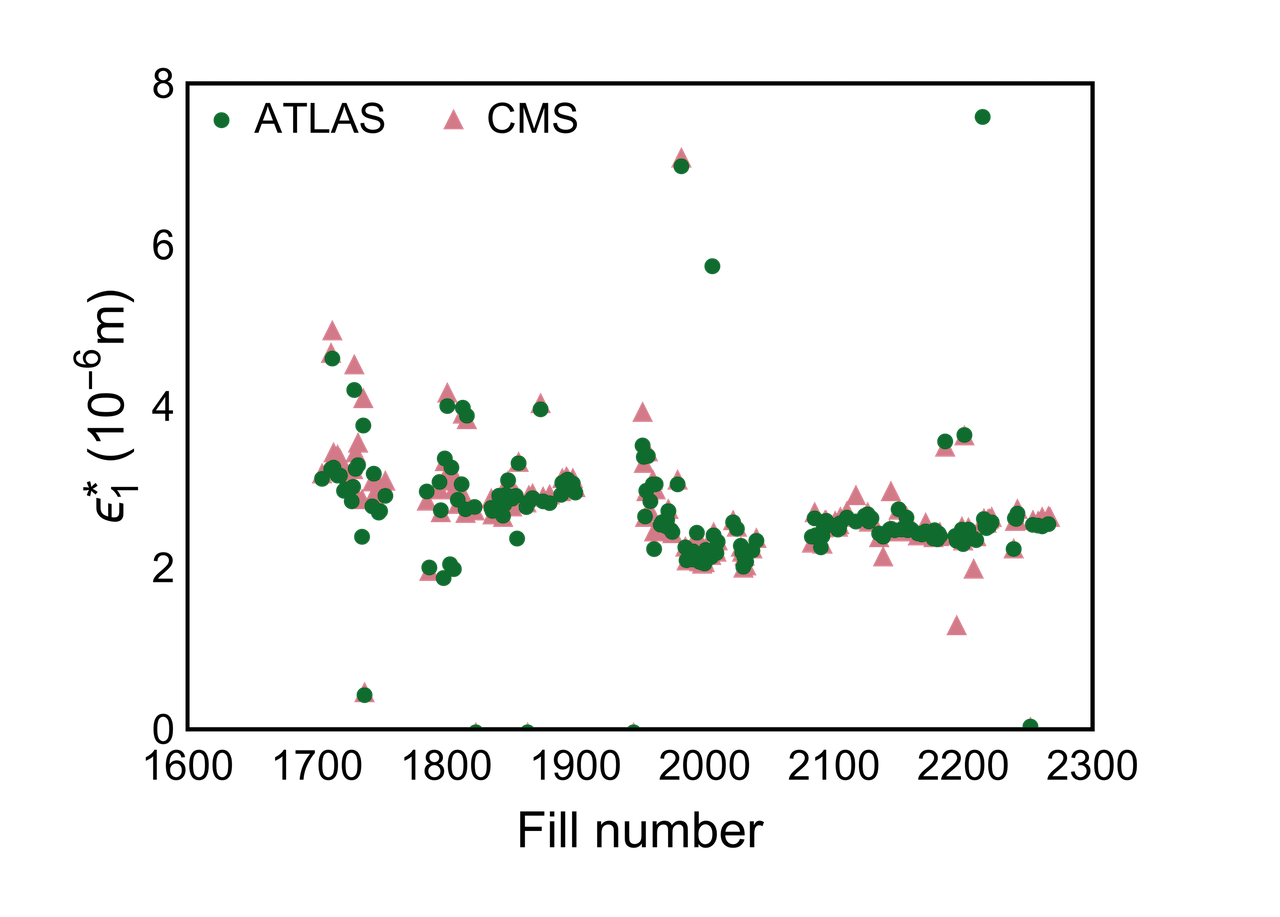} &
      \includegraphics[width=0.49\linewidth,clip=]{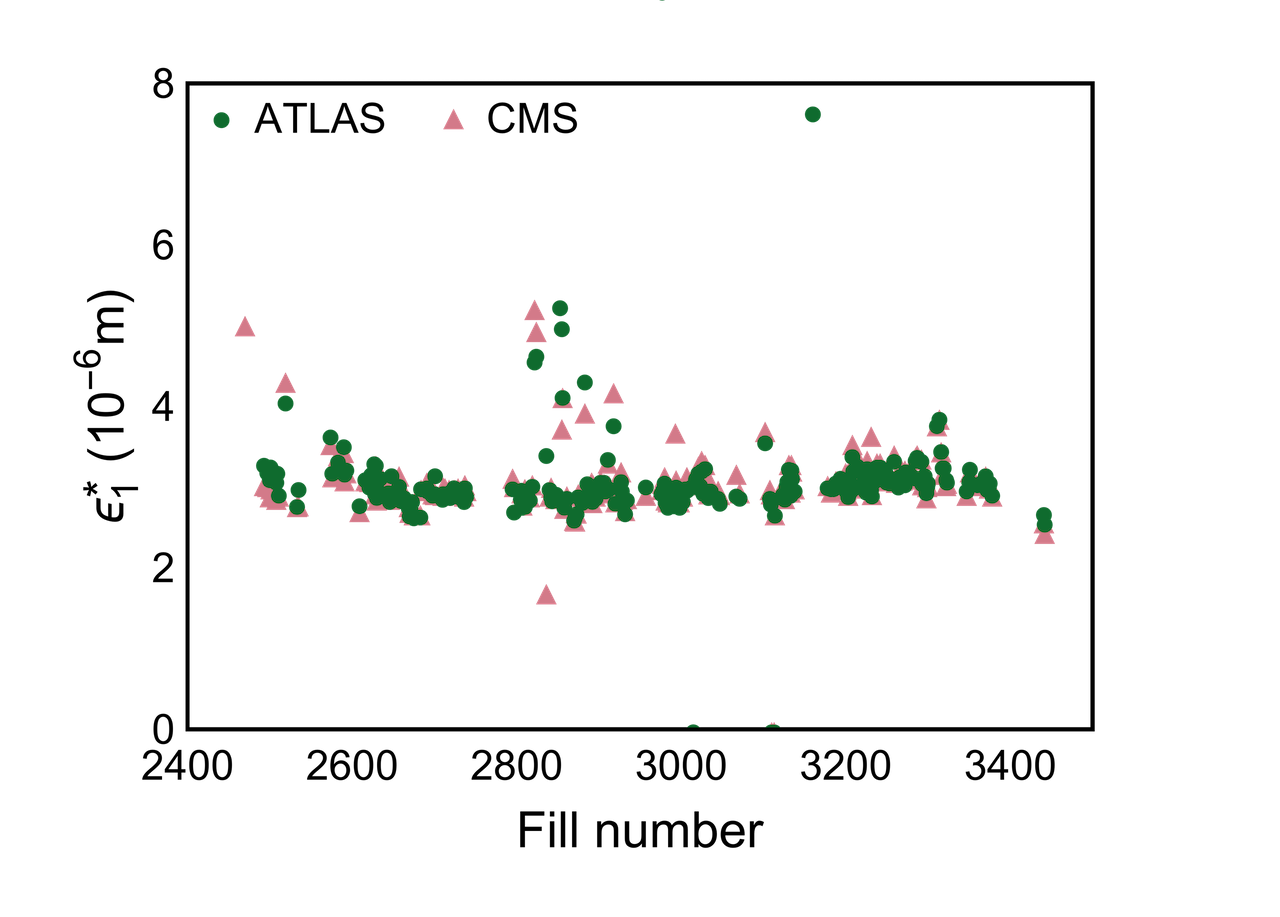} \\
      \includegraphics[width=0.49\linewidth,clip=]{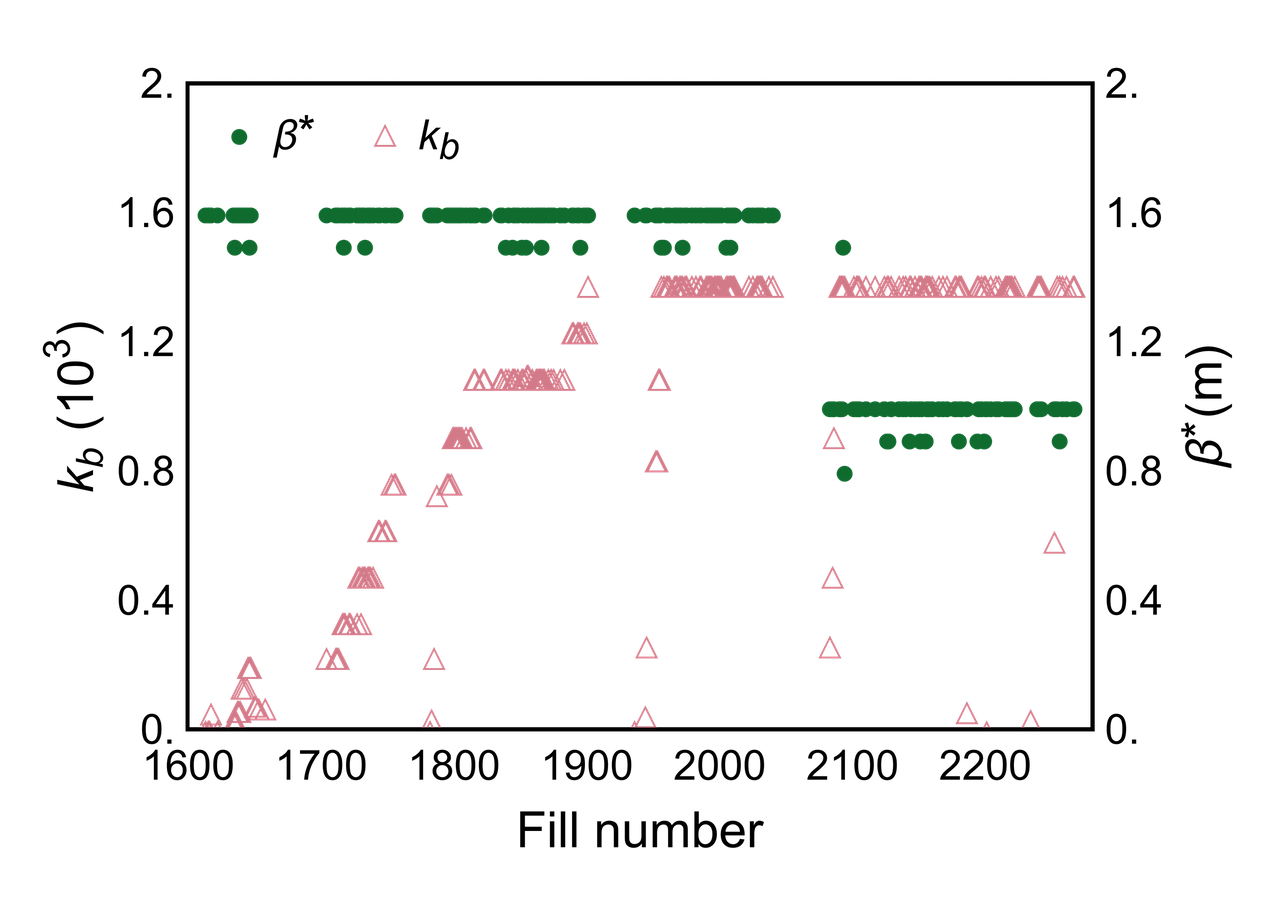} &
      \includegraphics[width=0.49\linewidth,clip=]{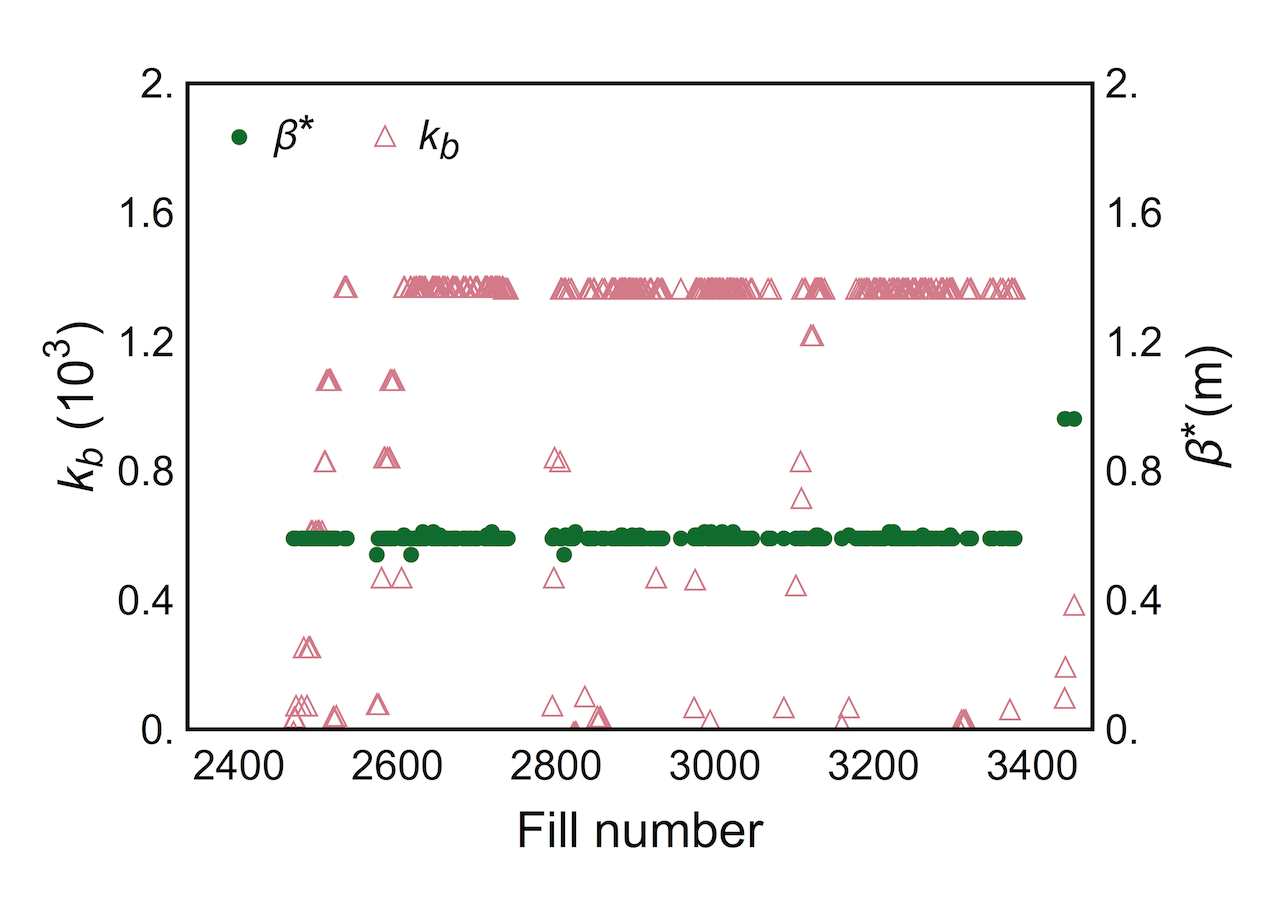} \\
    \end{tabular}
  \end{center}
  \caption{LHC performance during Run~1. The evolution of peak luminosity $L_{\rm i}$ and beam intensity (upper row), of $\epsilon^*_1$ (middle row) and of $\beta^*$ and $k_{\rm b}$ (lower row) is plotted vs. fill number. The data for 2011 (left) and 2012 (right) are shown for the sake of comparison. The estimate of $\epsilon^*_1$ assumes that the two beams have the same emittances  as well as the value of $\beta^*$ and is obtained by means of Eq.~\eqref{emitt_lumi}.}
  \label{summary_RunI}
\end{figure}

The models derived in the companion paper~\cite{lumi_Part_I} will be applied to the analysis of the LHC performance data collected during Run~1. Detailed information on this topic can be found in Refs.~\cite{RunI_1,RunI_2,RunI_3,RunI_4}, while in Ref.~\cite{MG_note} a preliminary analysis was made, without focusing on models to describe the luminosity and its time evolution. Here, the focus will be on the proton physics run and the data analysed can be found at~\cite{data_storage}. As an example, the evolution of some key parameters is shown in Fig.~\ref{summary_RunI} as a function of the fill number, which is an incremental integer number representing in a unique way the physics fill.
The peak luminosity $L$ and the total beam intensities are shown in the upper row of Fig.~\ref{summary_RunI}, while $\epsilon^*_1$ is reported in the middle row. The latter can be derived from the knowledge of $L$ and the beam parameters entering in Eqs.~\eqref{lumi} and~\eqref{geofac} according to 
\begin{equation}
 \epsilon^*_1 = \frac{- \theta_{\rm c}^2 \, \sigma_{z}^2 + 
\sqrt{\theta_{\rm c}^4 \, \sigma_{z}^4 +4 \, \chi^2}}%
{8 \, \beta^*}\qquad \chi = \frac{\gamma_{\rm r} \, f_{\rm rev} \, N_1 \, N_2 }%
{\pi \, k_{\rm b} \, L} \, ,
\label{emitt_lumi}
\end{equation}
with the assumption that the emittances of the two beams are equal as well as the value of $\beta^*$. In the lower row the evolution of $\beta^*$ and $k_{\rm b}$ is also shown. Data for the 2011 and 2012 runs are shown in the left and right columns, respectively. 

The total beam intensity has been increasing throughout the 2011 run, levelling out in 2012. Correspondingly, the peak luminosity has increased also because of the reduction of $\epsilon^*_1$ and $\beta^*$ and the increase of the number of bunches $k_{\rm b}$. A step decrease in $\epsilon^*_ 1$ is clearly seen in the 2011 data and corresponds to the progressive reduction of the controlled transverse emittance blow-up applied at the SPS~\cite{spsblowup}. Since that change, the emittance is fairly constant even during the 2012 run. It is worth mentioning that the beam brightness is defined by the LHC injectors' complex and is basically constant, which implies a linear relationship between intensity and transverse emittance. 

The evolution of $\beta^*$ follows a steady decrease, with a sudden jump during the 2011 run from $1.5$~m to $1$~m, whereas in 2012 it has been kept constant, but at the lower value of $0.6$~m. Finally, the number of bunches has been increased up to $1380$, corresponding to a bunch spacing of $50$~ns. During the 2011 run the gradual increase of $k_{\rm b}$ corresponding to the progress with the beam commissioning is clearly visible, while in 2012 the maximum number of bunches is the routine configuration. It is worth mentioning that the periods with reduced number of bunches correspond to the recommissioning after the regular technical stops occurring during the physics run. 

The data shown in Fig.~\ref{summary_RunI} are also used in the following analysis of the luminosity evolution. Among the full data set available from~\cite{data_storage} a selection has been considered including only the fills that resulted in successful physics runs, the so-called stable beams, of a total duration exceeding $10^3$~s and featuring $N_{\rm i,1,2} > 10^{13}$~p. Such a filtering allows removing data corresponding to beam commissioning stages or low luminosity fills, which would not be representative of the typical LHC performance. Additionally we only select those fills that have a number of bunches $k_b>1300$ to exclude ion runs. 

The analysis of this data set showed that the difference in beam intensity at the beginning of a physics fill is rather small, at the level of few percent~\cite{MG_note}. Hence, the simplifying assumption $N_{\rm i, 1}=N_{\rm i, 2}$ is fully justified and is used in the following. Using $L_{\rm i} = \Xi\, N_{\rm i,1}\, N_{\rm i,2}$ we can calculate $\varepsilon$ from the initial luminosity $L_{\rm i}$:
\begin{equation}\label{epsilon}
\varepsilon =
	\frac{\sigma_{\rm int}\, n_c\, L_{\rm i}}{f_{\rm rev} \,N_{\rm i,1} N_{\rm i,2}}
\, ,
\end{equation}
where $n_c=2$ because the vast majority of protons are burnt in the two high-luminosity interaction points (see the previous comment on the relative luminosities of the various LHC experiments), and the total inelastic cross-section for proton-proton collisions is $\sigma_{\rm int}$ is $73.5$~mb for $3.5$~TeV and $76$~mb for $4$~TeV~\cite{inel1,inel2} for protons, representing the total inelastic cross-section.
\subsection{Observed time-dependence of beam parameters and other assumptions for data analysis}\label{sec:time-dep}
The analyses presented in Ref.~\cite{lumi_Part_I} included the situation when some beam parameters are changing during the fill, as it can be the case for the rms bunch length $\sigma_z$ or any of the two transverse normalised emittances $\epsilon^*_{x,y}$. 

Equations~\eqref{lumi} and~\eqref{geofac} show that while $\sigma_z$ has an impact on $F$, only, the transverse normalised emittances $\epsilon^*_{x,y}$ affect both $F$ and the peak luminosity.

The measured data revealed that the variation of $\sigma_z$ over a typical physics fill does not exceed $\approx 7~\%$. Such a small variation is understandable by considering that at the collision energy of $3.5$~TeV or $4$~TeV, which are the values for the 2011 and 2012 runs, respectively, the damping generated by synchrotron radiation is not too strong. Therefore, the time-dependence of $\sigma_z$ can be safely neglected in the analyses presented in the following sections. 

\begin{figure}[htb]
  \begin{center}
        \includegraphics[width=0.59\linewidth,clip=]{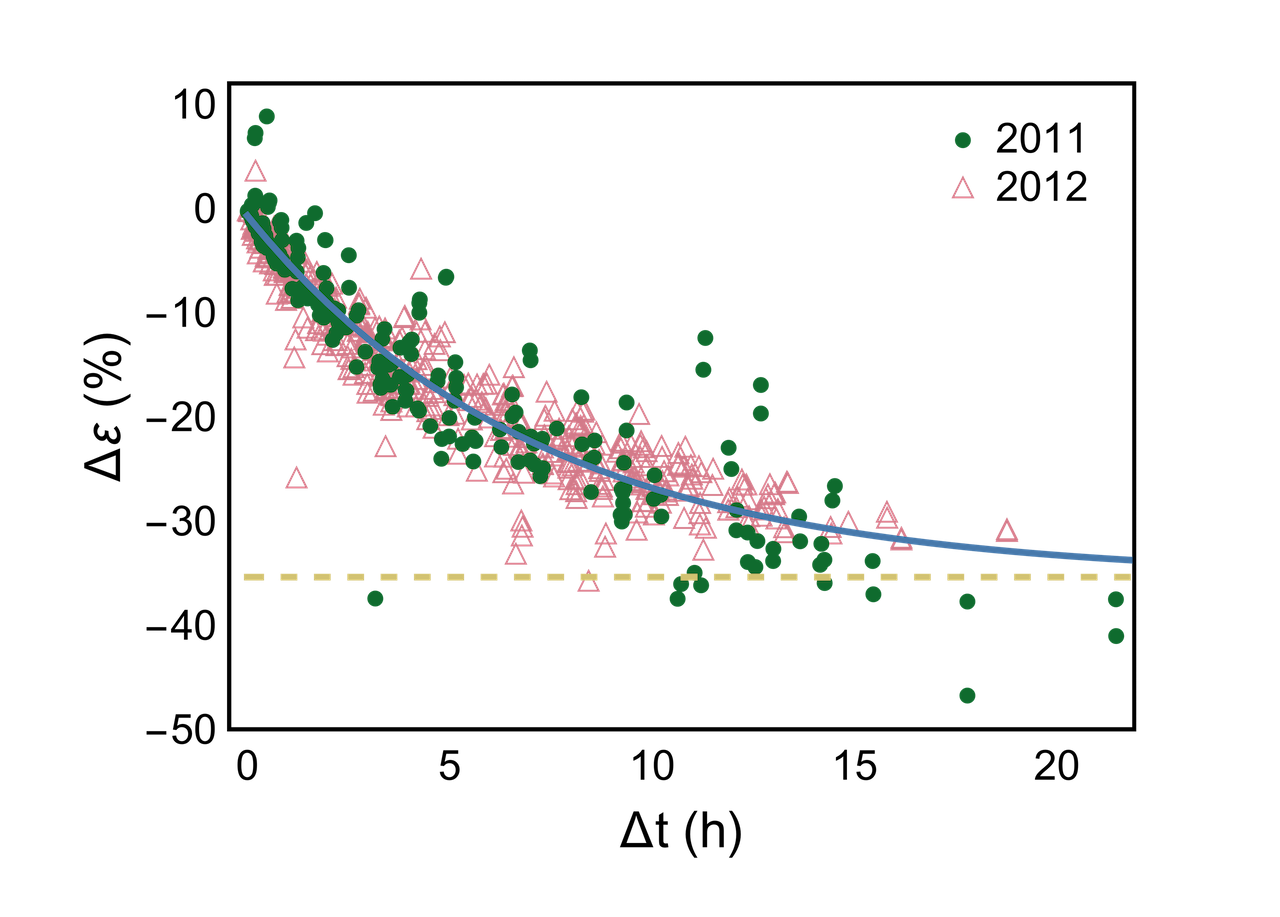}
  \end{center}
  \caption{Relative difference of $\varepsilon$ between end and beginning of a physics fill as a function of the fill length for the selected data from the 2011 and 2012 runs. The continuous  curve represents an exponential fit to the data, while the dashed one the asymptotic value.} 
  \label{time-dependence}
\end{figure}

The time-dependence of $\varepsilon$ needs to be assessed to decide the approach to be applied to the data analyses and the outcome of these investigations, based on the selected fills of the 2011 and 2012 runs, is shown in Fig.~\ref{time-dependence}. 

The data have been fitted using an exponential function and the result is given by 
\begin{equation}
\Delta \, \varepsilon (t) = 34.69 \, \Exp{-0.1358 \, t} - 35.39
\end{equation}
where $t$ is expressed in hours and $\Delta\varepsilon$ in percent. The fit quality is given by $R_\text{adj}^2 = 96.17$ (see Eq.~\eqref{eq:adj} in section~\ref{sec:lumi_data} for the definition of this quantity). For the majority of fills $\Delta \, \varepsilon$ does not exceed $\approx 30~\%$ and it has been decided to perform the numerical analyses assuming a time-independent $\varepsilon$. Note that the fit models presented in section~\ref{sec:lumi_data} have been cross-checked against models in which $\varepsilon$ had been assumed to be time-dependent and the differences have been found small, thus confirming that the assumption made is appropriate. It is worth mentioning that it is planned to apply the most general model described in Ref.~\cite{lumi_Part_I} to the description of data for the LHC Run~2~\cite{lumi_Part_III}.

Another aspect to consider is whether some of the parameters entering in the proposed models should be different for the two beams. In fact, in Ref.~\cite{lumi_Part_I} the descriptive models could include beam-specific values for both the initial beam intensity and the pseudo-diffusive effects. A close inspection of the Run~1 data~\cite{MG_note} reveals that for a typical physics fill the quantity $2 \, |N_{\rm i,1}-N_{\rm i,2}|/(N_{\rm i,1}+N_{\rm i,2})$ does not exceed $\approx 10~\%$. Hence, in the analysis reported in the following sections, the two initial beam intensities have always been assumed equal and the corresponding model for the burn-off part has been used~\cite{lumi_Part_I}. Given that a similar estimate holds also for the intensities at the end of the physics fills, the pseudo-diffusive effects have been assumed to be the same for both beams.
\section{Time evolution of luminosity over a fill}\label{sec:lumi_data}
The first step in the analysis of the LHC Run~1 data is the fit of the pseudo-diffusive component of the luminosity evolution based on the expression, which was derived in~\cite{lumi_Part_I}, given by 
\begin{align}
\frac{L(\tau)}{L_{\rm i}} & = \displaystyle{\frac{1}{\left [ 1 + \varepsilon \, N_{\rm i} \, (\tau-1) \right ]^2}}  + \\ \label{Lpd unint}
& \phantom{=} -  \displaystyle{\left [ \Exp{-\frac{D^2(\tau)}{2}} - \Exp{-\frac{D^2(1)}{2}}\right ] \left \{ 2- \left [ \Exp{-\frac{D^2(\tau)}{2}} - \Exp{-\frac{D^2(1)}{2}}\right ] \right \}} \notag \, ,
\end{align}
where $L_{\rm i}$ is the initial value of the luminosity, given by $L_{\rm i}=\Xi \, N^2_{\rm i}$, and $N_{\rm i}$ the corresponding initial value of the beam intensity. 

\begin{figure}[htb]
  \begin{center}
    \begin{tabular}{@{}c@{}@{}c@{}}
      \includegraphics[width=0.49\linewidth,clip=]{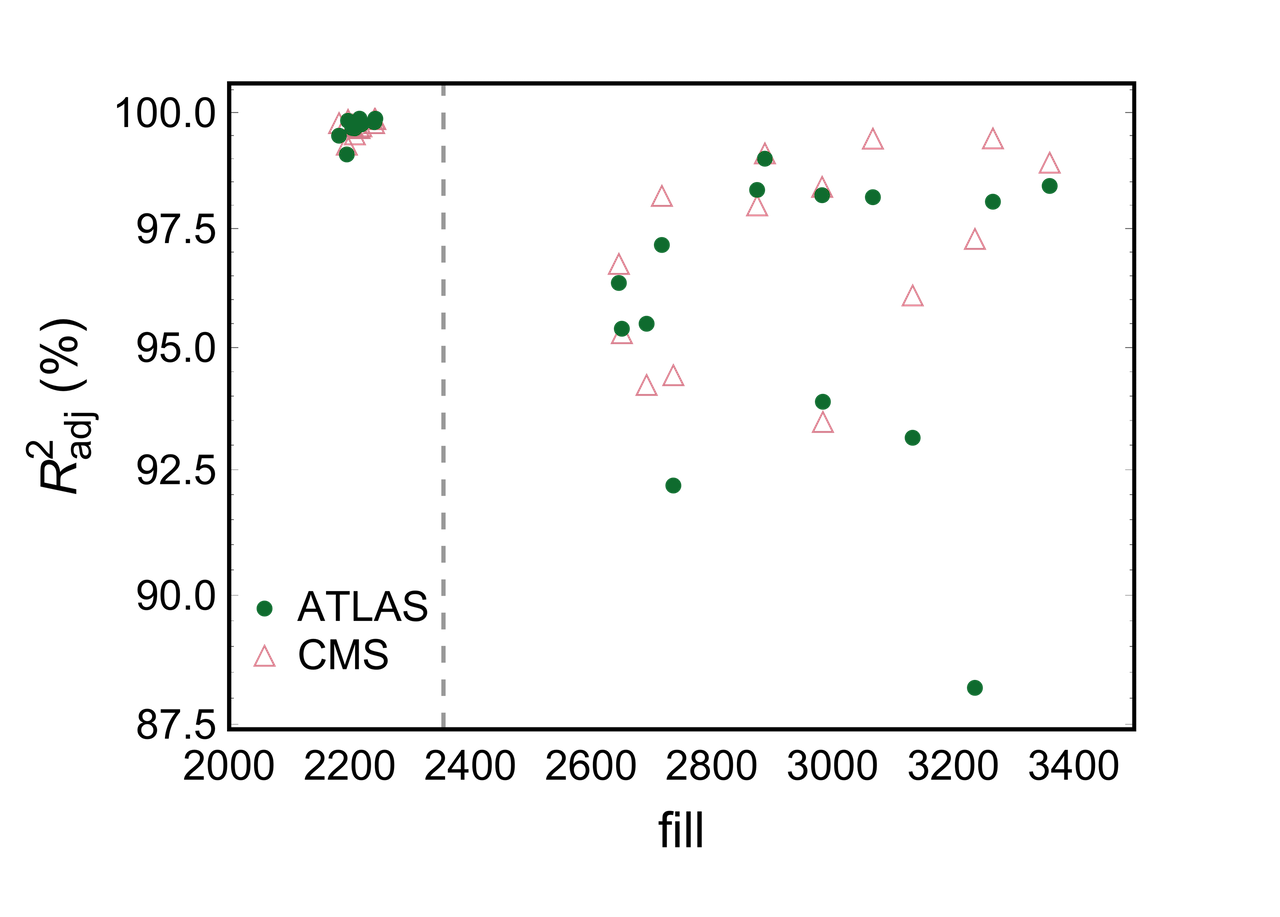} &
      \includegraphics[width=0.49\linewidth,clip=]{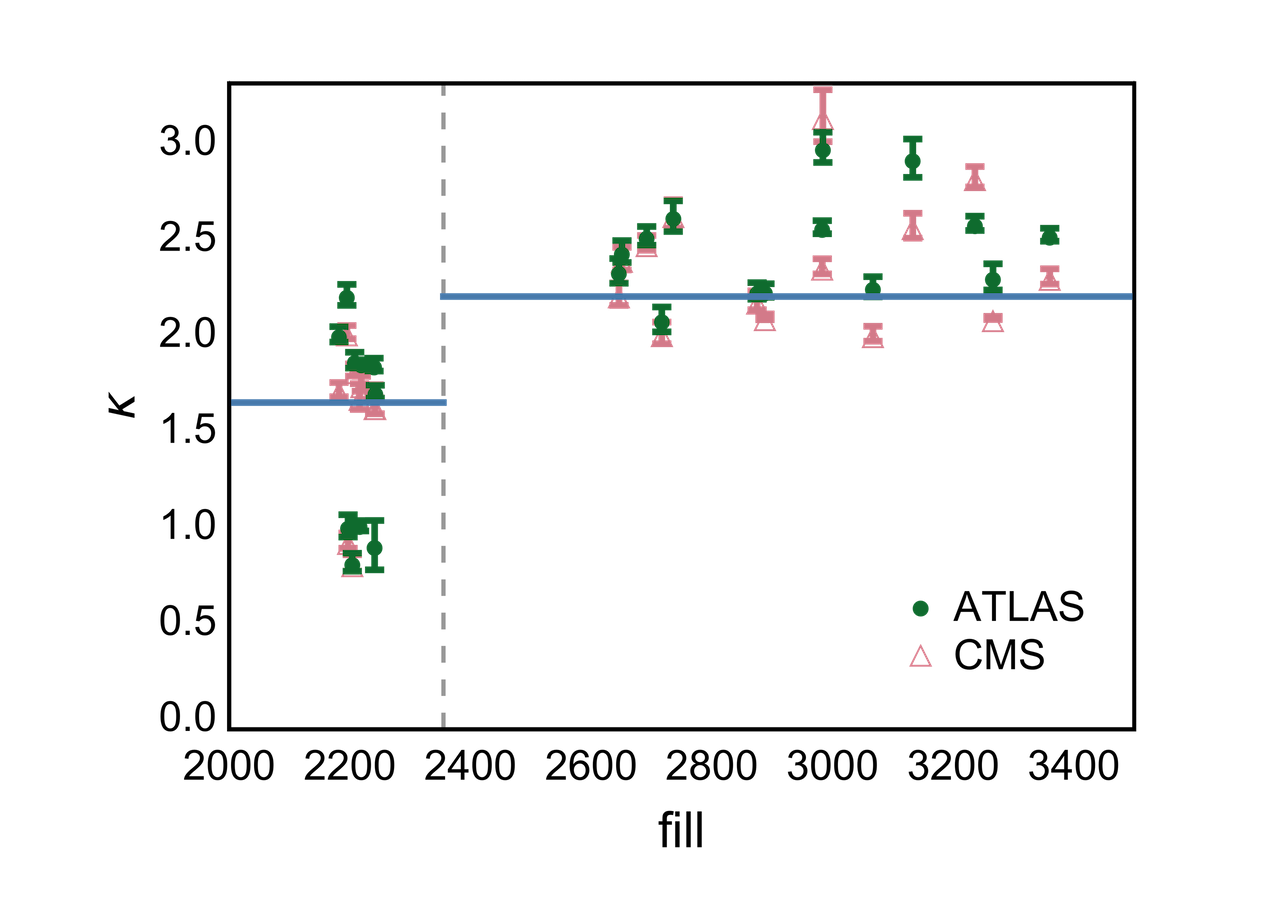} \\
      \includegraphics[width=0.49\linewidth,clip=]{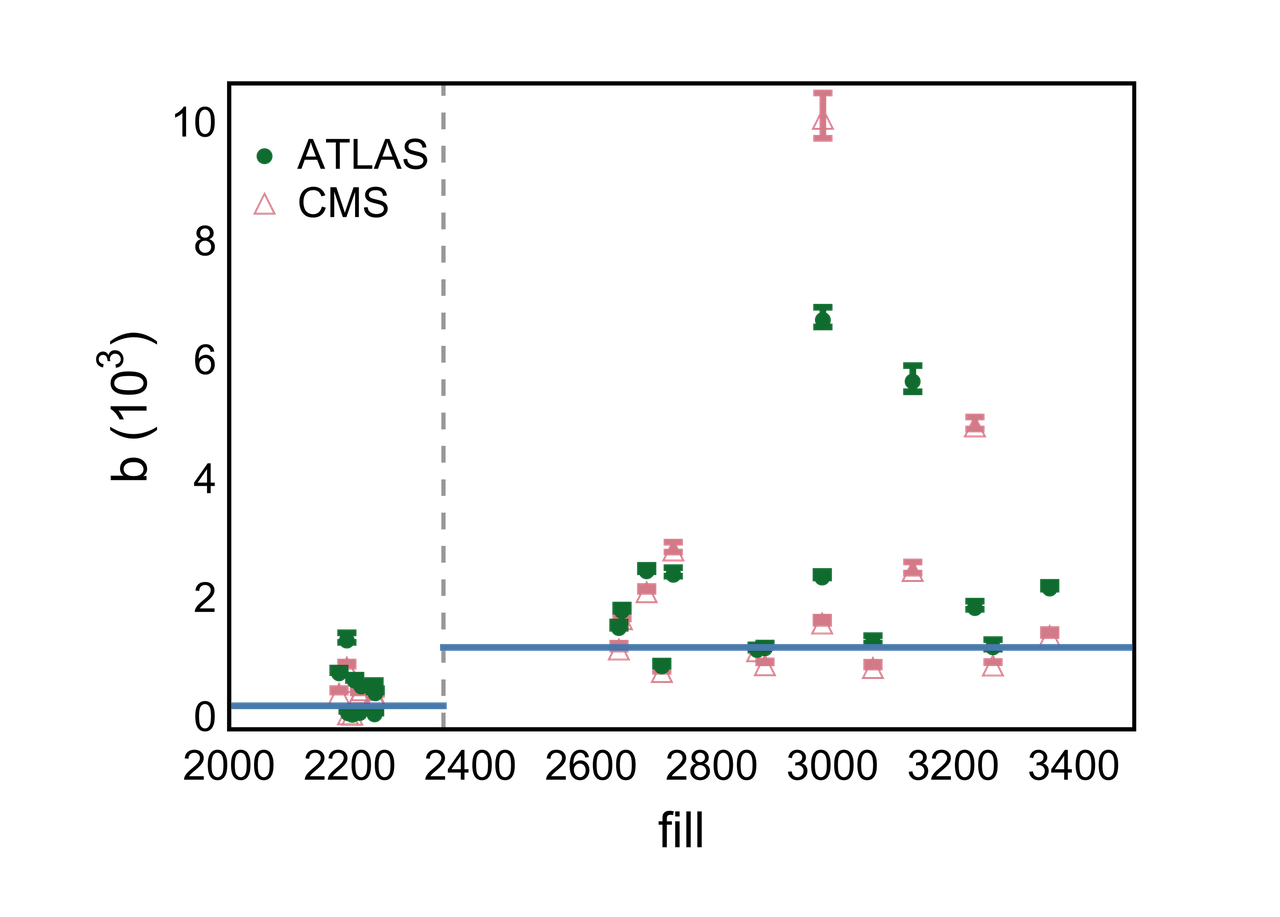} &
      \includegraphics[width=0.49\linewidth,clip=]{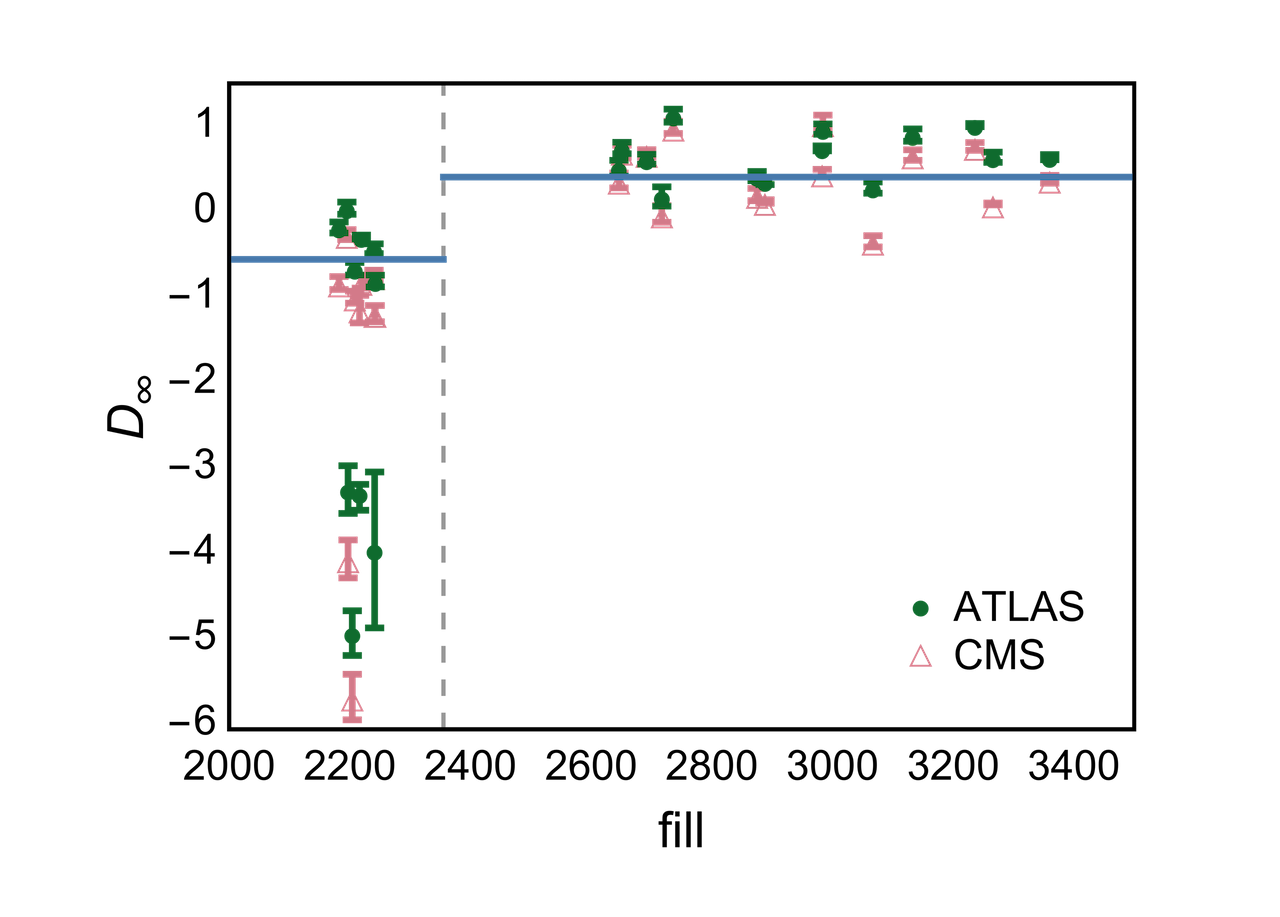} \\
      \includegraphics[width=0.49\linewidth,clip=]{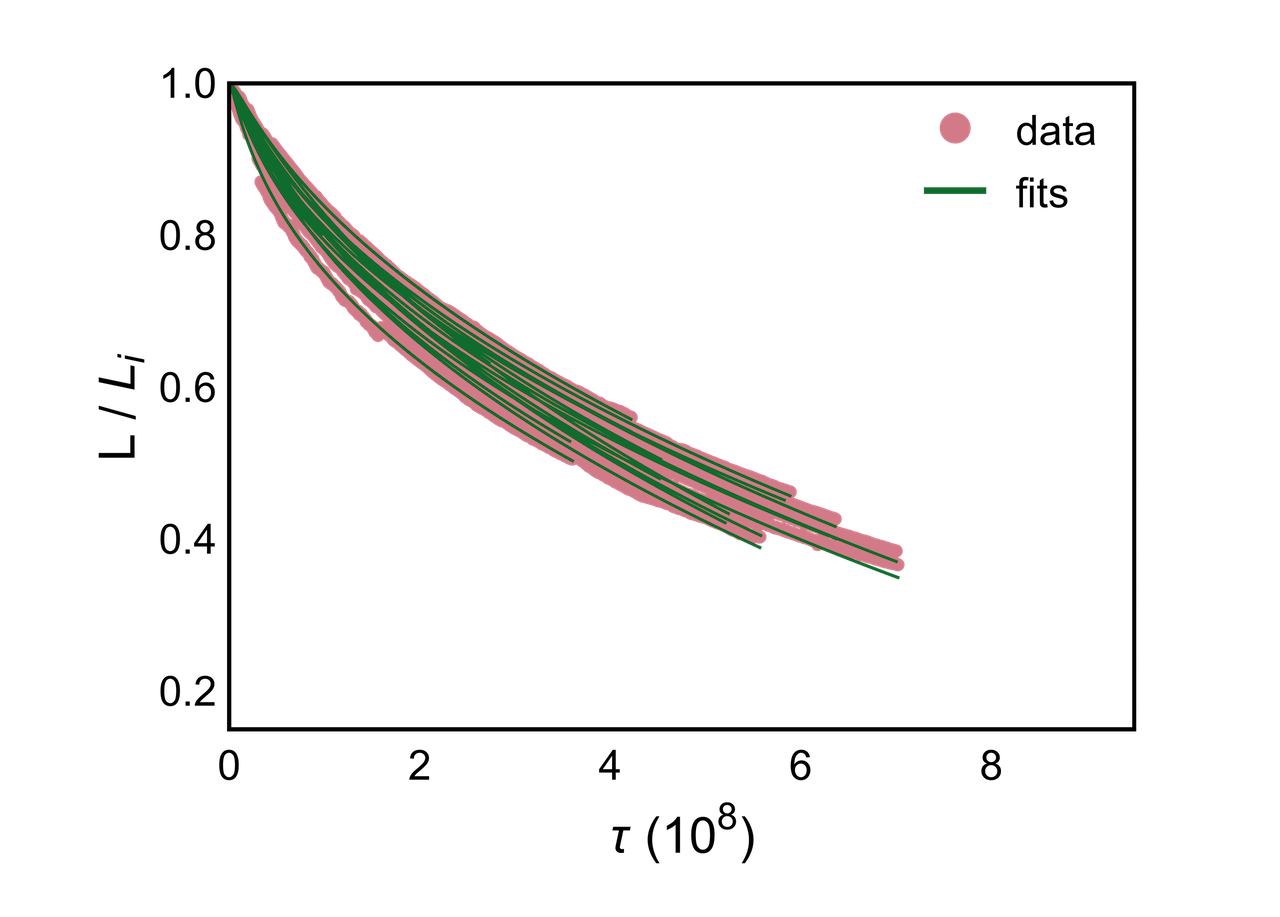} &
      \includegraphics[width=0.49\linewidth,clip=]{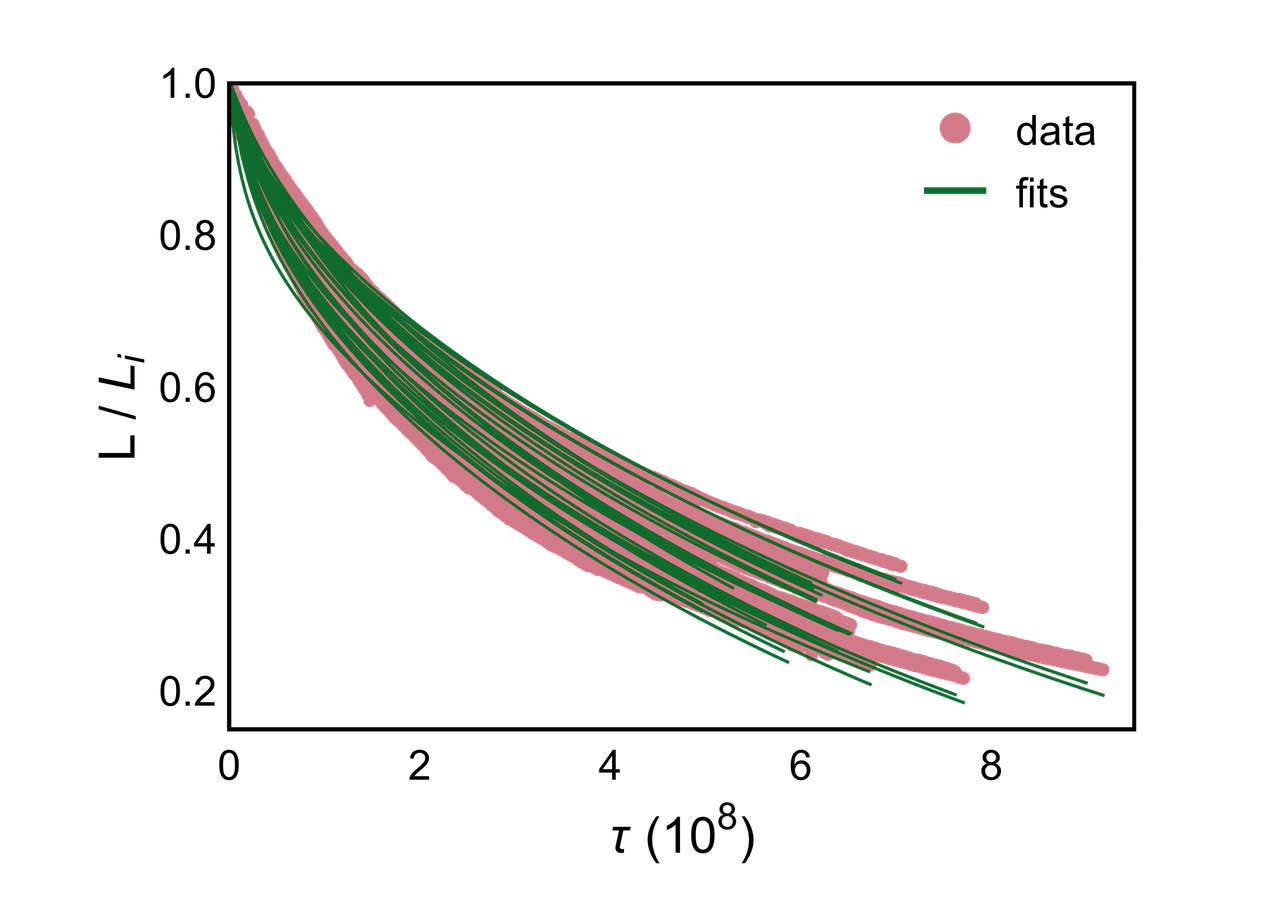}
    \end{tabular}
  \end{center}
  \caption{Fit parameters and $R^2_\text{adj}$ for the pseudo-diffusive contribution $L^{\rm pd}(\tau)$ per fill for the case of a fit performed with three free parameters. The vertical dashed line separates the fills of 2011 from those of 2012. The error bars on the parameters are estimated using the BCa method (see \ref{app:difficulties}). The horizontal line represents the weighted (over the fit parameter's error) average per year. Note the clear distinction between the results for the two years. The two bottom plots show the measured and fitted curves for $L$ (normalised to the initial fill luminosity $L_{\rm i}$) for 2011 (left) and 2012 (right) fills and a very good agreement is clearly visible.} 
  \label{unintegrated}
\end{figure}

For this, 24 fills, 10 from 2011 and 14 from 2012, have been selected and fitted individually, also separating the results for the two high-luminosity experiments, ATLAS and CMS. The results are shown in Fig.~\ref{unintegrated} where the model parameters have been reported as a function of the fill number and their errors are estimated using the BCa method (see \ref{app:difficulties}). Also shown is $R^2_\text{adj}$, the so-called adjusted coefficient of determination given by
\begin{equation}
R^2_\text{adj} =
	1 - \frac{N-1}{N-\nu-1}\;\frac{\Sigma^2}{\sigma^2} \, ,
    \label{eq:adj}
\end{equation} 
where $N$ is the sample size, $\nu$ the number of fit parameters, $\Sigma^2$ the sum of residues squared defined in Eq.~\eqref{eq:RSS}, and 
\begin{equation}
	\sigma^2 = \sum\limits_{i=1}^N \left [ y_i-\bar{y}\right]^2
\end{equation}
is the total variance of the data with $\bar{y}$ the average over all $y_i$. Note that $R^2_\text{adj}$ compares the fit under consideration to the simplest fit, i.e. a constant line through the mean. When $R^2_\text{adj}\ll 1$ (or possibly even negative), the fit is of poor quality as the mean of the data provides a better fit than the proposed model. A good fit has $R^2_\text{adj}\to 1$, indicating that the residues are small compared to the data variance.

If we look at the results in Fig.~\ref{unintegrated}, we notice that all fits are of particular good quality, as all except one have $R^2_\text{adj} > 90~\%$, while for all fits from 2011 this is even $R^2_\text{adj} > 99~\%$. There is a clear distinction between the results for 2011 and those for 2012, both in spread, but also in behaviour, as the yearly average value of $D_\infty$ is negative for 2011 whereas it is positive for 2012. The larger spread in the fit parameters for the 2012 run might be generated by the transverse instabilities that plagued that run~\cite{Inst2012}. It is interesting to stress that from Fig.~\ref{unintegrated} no clear systematic difference between the fitted models based on the ATLAS or CMS data is found as the variations change on a fill-by-fill basis. The differences observed are not always within the error bars, which might be underestimated. Moreover, some differences are to be expected as $\beta^*$ waist position correction was not fully mastered, the first being at the level of $5~\%$~\cite{OMC1}. A further refinement has been applied during Run~2 lowering the uncertainty on $\beta^*$ at the level of about $1~\%$~\cite{OMC2}.

Following the discussion in \ref{app:difficulties}, it is useful to fit the data to a slightly adapted model, which has a reduced set of parameters. To this end, we selected three different configurations: one where we fix $\kappa=2$ (according to the Nekhoroshev estimate \cite{nekhor}) and fit $b$ and $D_\infty$; one where we fix $D_\infty=0$ (as it is approximately the average of Run 1) and fit $b$ and $\kappa$; and one where we fix both $\kappa=2$ and $D_\infty=0$ and fit only $b$, thus leaving only one model parameter.

\begin{table}[htb]
\centering
\caption{Summary of the fit parameters and associated errors corresponding to the expression of $L^{\rm pd}(\tau)$, for different model parameters and for different data subsets. The parameter values presented are the weighted (over the fit parameter's error) averages over the fills, whereas a regular average is used for $\bar{R}^2_\text{adj}$.}
\begin{tabular}{@{}l@{}cccc@{}}
\hline
& $D_\infty$ & $b$ & $\kappa$ & $\bar{R}^2_\text{adj} [\%]$ \\ \hline
2011 data \qquad &
$ -0.61 \pm 0.71 $ & $ 180 \pm 210 $ & $ 1.64 \pm 0.40 $ & 99.759\\
2011 data, $\kappa = 2$\qquad &
$ -0.44 \pm 0.19 $ & $ 920 \pm 73 $ & -- & 99.736\\
2011 data, $D_\infty = 0$\qquad &
-- & $ 1900 \pm 940	$ & $ 2.41 \pm 0.19 $ & 99.726\\
2011 data, $\kappa = 2$, $D_\infty = 0$\qquad &
-- & $ 752 \pm 18 $ & -- & 97.469\\ \hline
2012 data \qquad &
$ 0.36 \pm 0.41 $ & $ 1200 \pm 680 $ & $ 2.19 \pm 0.24 $ & 96.531\\
2012 data, $\kappa = 2$\qquad &
$ 0.20 \pm 0.26$ & $ 670 \pm 110 $ & -- & 96.232\\
2012 data, $D_\infty = 0$\qquad &
-- & $ 200 \pm 200 $ & $ 1.84 \pm 0.26 $ & 96.037\\
2012 data, $\kappa = 2$, $D_\infty = 0$\qquad &
-- & $ 748 \pm 23 $ & -- & 93.492 \\
\hline
\label{fit_par_unint}
\end{tabular}
\end{table}

The resulting weighted average parameter values are listed in Table~\ref{fit_par_unint}, where the associated error is given by the weighted standard deviation and $\bar{R}^2_\text{adj}$ is the average over the fills. The difference between the fills from 2011 and from 2012 persists in all four fit versions, for this reason we did not calculate the total average parameter values over the two years of Run~1. Note that when one parameter is fixed ($\kappa=2$ or $D_\infty=0$) the fit quality is almost unaffected (having only a slight decrease), but when two parameters are fixed (both $\kappa=2$ and $D_\infty=0$ at the same time), there is a clear worsening of the fit (even though the overall quality remains rather good). This confirms the considerations reported in \ref{app:difficulties} that fixing one parameter delivers a fit that is as good as using the full parameter set, given the existence of an approximate degeneracy of the parameter space. The case $\kappa=2$ is preferred over $D_\infty=0$, because of its justification on the basis of the Nekhoroshev theorem.
\section{Integrated luminosity over a physics fill}\label{sec:int_lumi_data}
The second step consists of establishing the model for the integrated luminosity delivered in a single fill for physics. In Ref.~\cite{lumi_Part_I} it has been shown that, under the assumption of considering burn-off phenomena only, the integrated luminosity over a fill can be expressed as 
\begin{equation}
L_{\rm norm}^\text{bo}(\bar{\tau})=
	\frac{\bar{\tau}}{1+\bar{\tau}}
\label{absv1_1}
\end{equation}
with an appropriate rescaling and by using a normalised time given by $\bar{\tau}=\varepsilon \, N_{\rm i} \, (\tau-1)$. Whenever pseudo-diffusive effects are taken into account, then one can assume the following form for the luminosity evolution
\begin{equation}
L_{\rm norm}(\tau) = L_{\rm norm}^{\rm bo}(\tau) +  L^\text{pd}(\tau)
\label{scaling1} 
\end{equation}
where $L_{\rm norm}^{\rm bo}$ stands for the burn off component of the luminosity evolution~\cite{lumi_Part_I} and $L^\text{pd}$ can be expressed at first order in the small parameter $\varepsilon$ as
\begin{equation}\label{Lpdfitmodel}
L^{\rm pd}(\tau)=
  - N_\text{i}\,\varepsilon\Int{1}{\tau}{\tilde{\tau}}
    \left[\Exp{-\frac{D^2(\tilde{\tau})}{2}} - \Exp{-\frac{D^2(1)}{2}}\right]
    \left\{
      2 - \left[ \Exp{-\frac{D^2(\tilde{\tau})}{2}} - \Exp{-\frac{D^2(1)}{2}}\right] 
    \right \}
\end{equation}
where $N_{\rm i}$ is the initial intensity.

\begin{figure}[htb]
  \begin{center}
    \begin{tabular}{@{}c@{}@{}c@{}}
      \includegraphics[width=0.49\linewidth,clip=]{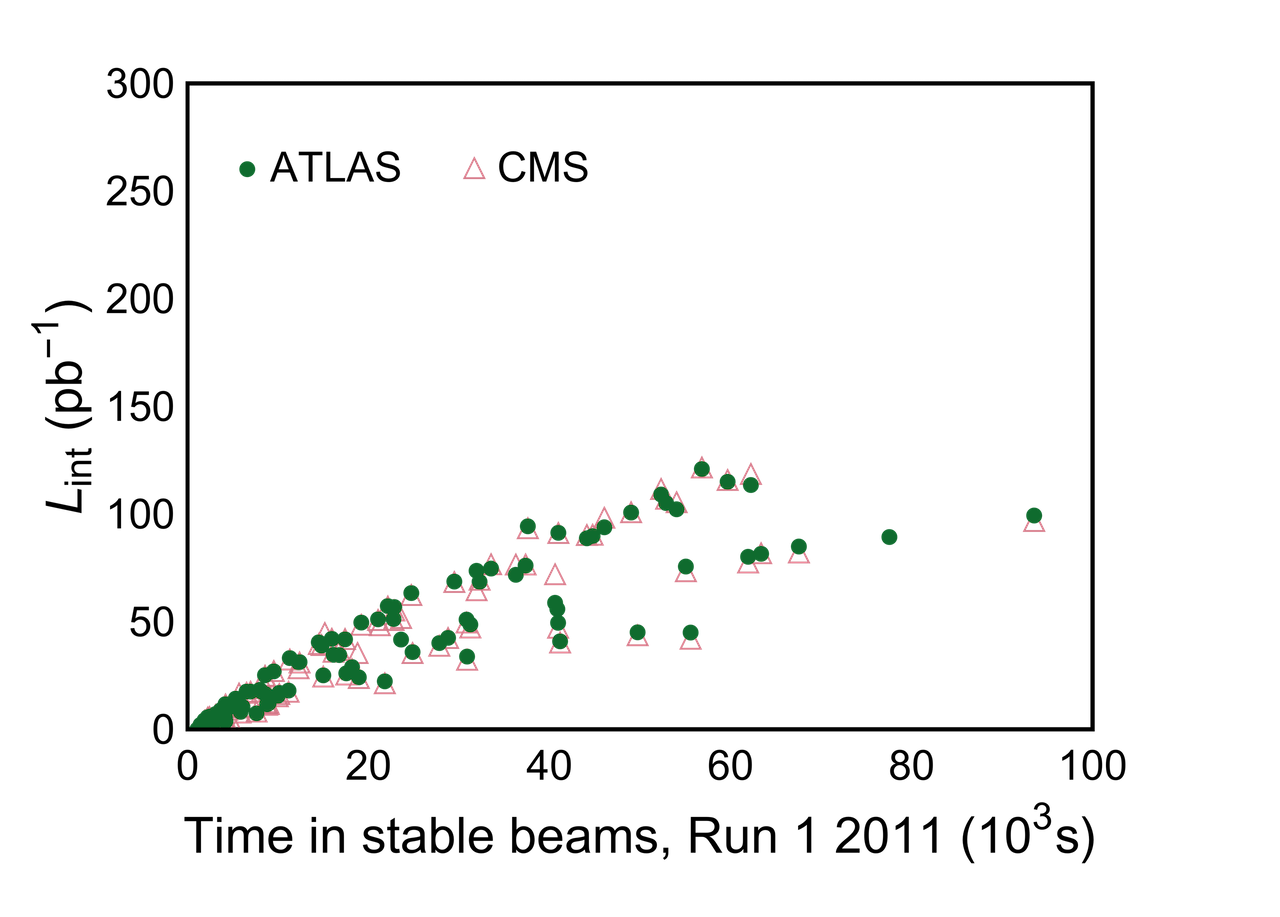} &
      \includegraphics[width=0.49\linewidth,clip=]{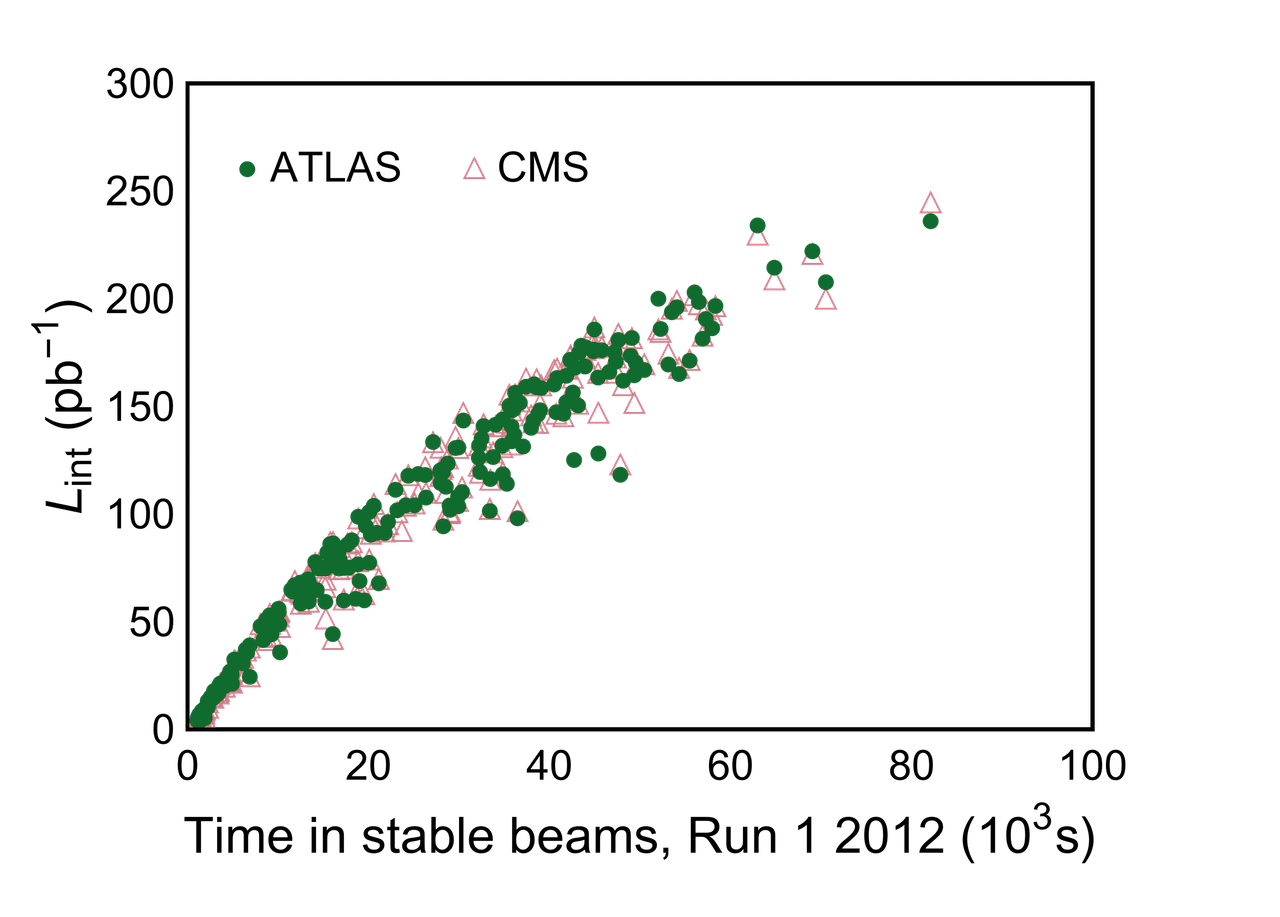}\\
      \includegraphics[width=0.49\linewidth,clip=]{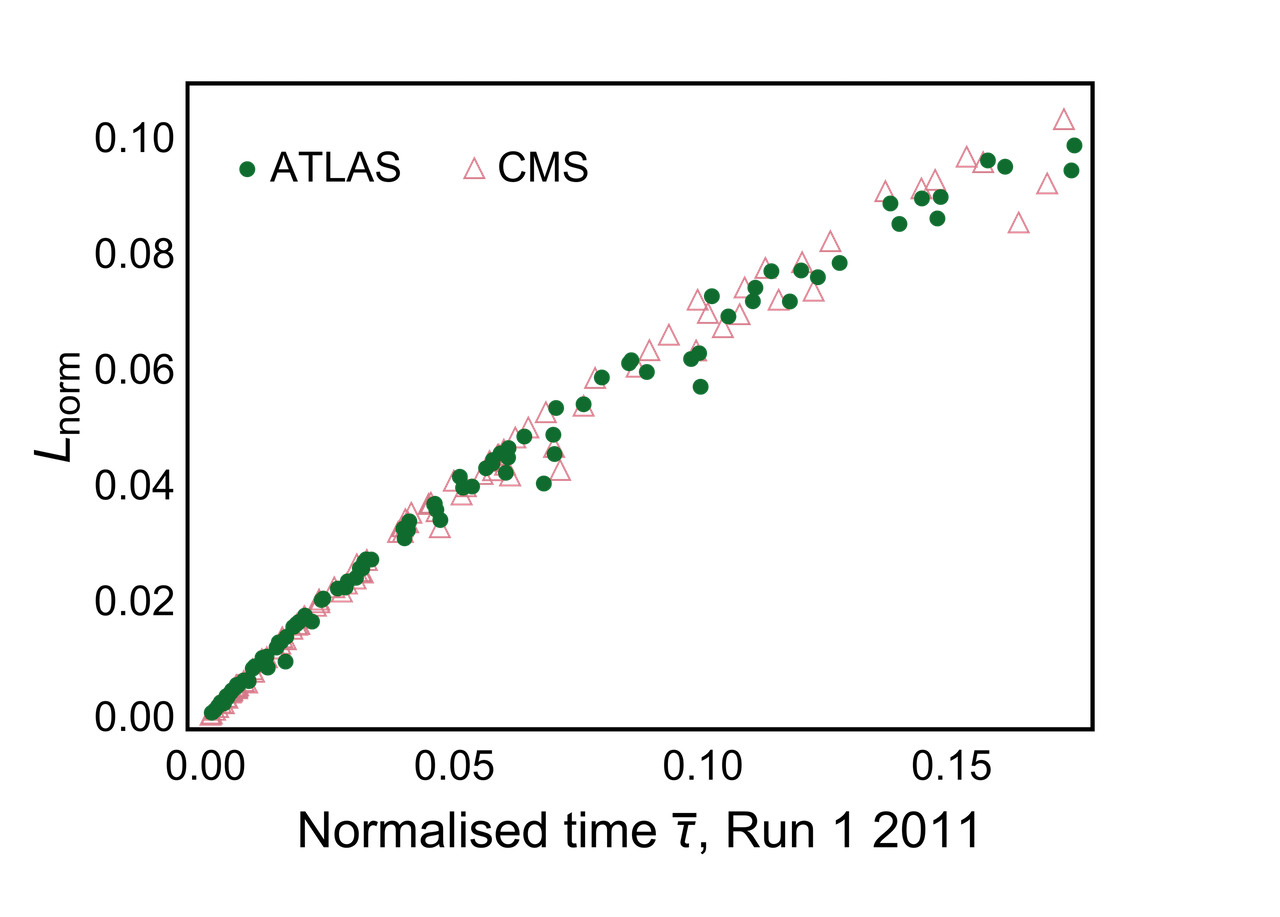} &
      \includegraphics[width=0.49\linewidth,clip=]{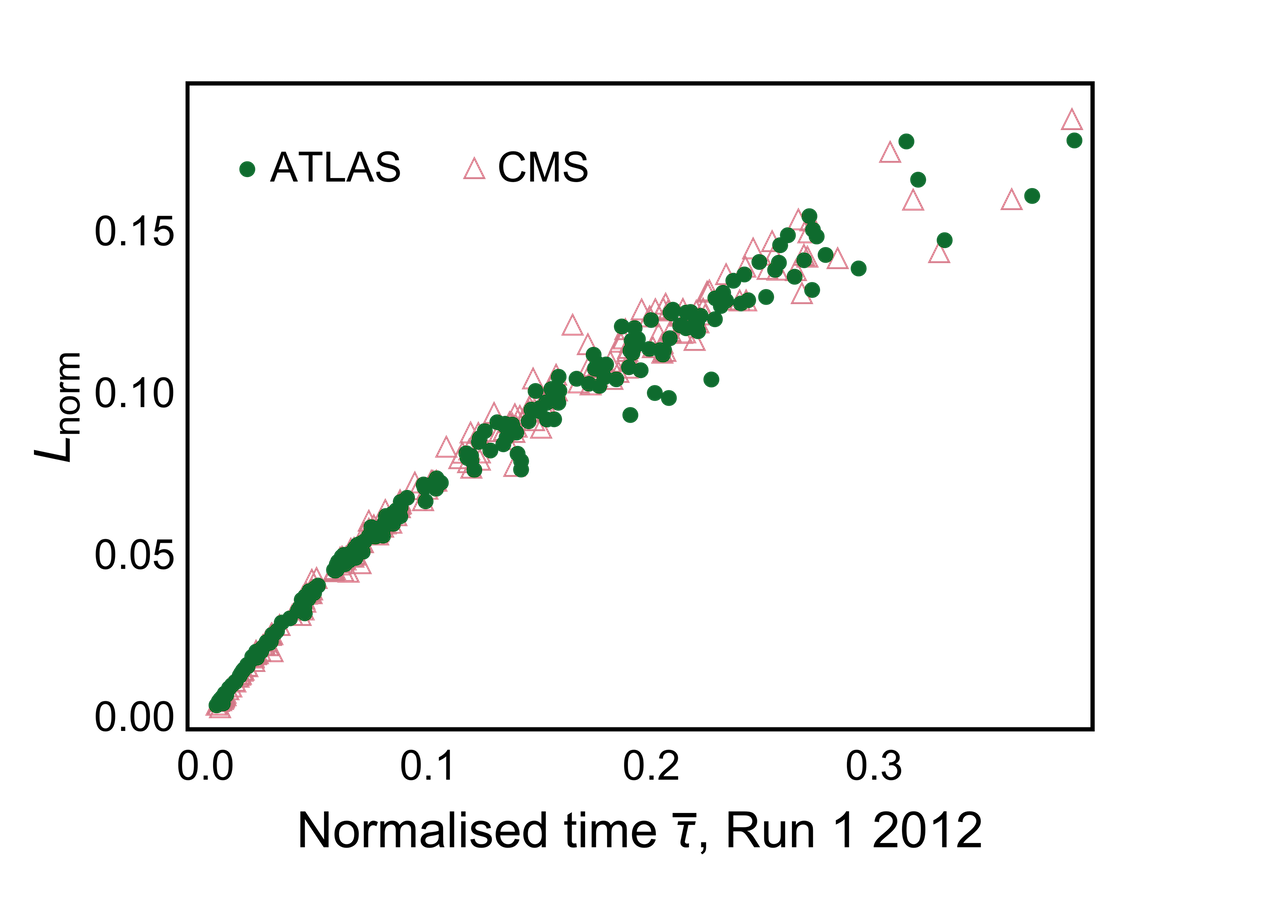}\\
    \end{tabular}
  \end{center}
  \caption{The integrated luminosity delivered in a single fill for physics is shown as a function of the fill duration (upper row). The normalised integrated luminosity defined according to Eq.~\eqref{scaling1} is also plotted as a function of the normalised time (lower row). The data for both 2011 (left) and 2012 (right) runs are shown.}
  \label{intlumi}
\end{figure}

The plot of $L_{\rm int}$ (the integrated luminosity over a physics fill) is shown in Fig.~\ref{intlumi} (upper row) for 2011 (left) and 2012 (right) runs. 

The large spread observed for the 2011 data is due to the change of beam and ring parameters, i.e. transverse emittance, intensity, and $\beta^*$ occurred during the year 2011 (see Fig.~\ref{summary_RunI}), whereas the situation in terms of beam parameters has been much more stable throughout the 2012 run. The impact of the proposed normalisation of both $L_{\rm int}$ and $\tau$ according to Eq.~\eqref{absv1_1} is also shown in Fig.~\ref{intlumi} (lower). The spread of the data points is almost completely removed and a sort of universal curve is appearing, with a similar shape for both 2011 and 2012 data. The normalised luminosity allows for an easy recognition of outlying data points, which have been removed (in total 12 data points) when fitting the data to the evolution model.

The luminosity data are given as a function of time instead of number of turns, hence the least computationally expensive way to obtain the pseudo-diffusive contribution to luminosity from these data is to rearrange Eq.~\eqref{scaling1} into
\begin{equation}\label{Lpdcalculation}
L^{\rm pd}(\tau) =
	\varepsilon N_{\rm i,max}\;f_{\rm rev} \frac{\hat{L}_{\rm int}(t)}{L_{\rm i}}
  -L_{\rm norm}^{\rm bo}(\tau)
\end{equation}
and use Eq.~\eqref{epsilon} to evaluate $\varepsilon$ for every data point. When the data, calculated from Eq.~\eqref{Lpdcalculation}, has been fitted to the model~\eqref{Lpdfitmodel}, it is recast into the normalised luminosity using Eq.~\eqref{scaling1}.

\begin{figure}[htb]
  \begin{center}
     \begin{tabular}{@{}c@{}@{}c@{}}
      \includegraphics[width=0.49\linewidth,clip=]{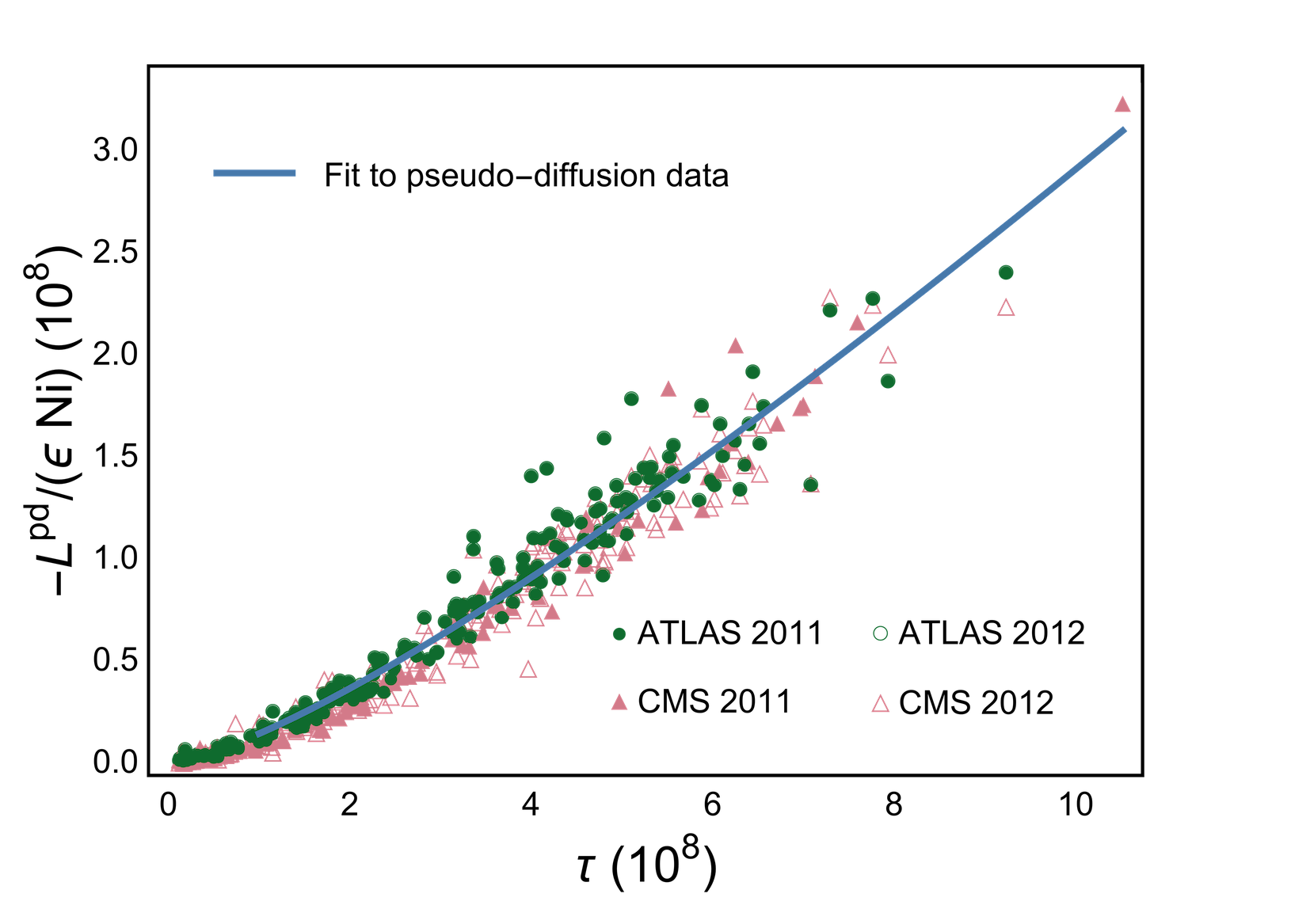} &
      \includegraphics[width=0.49\linewidth,clip=]{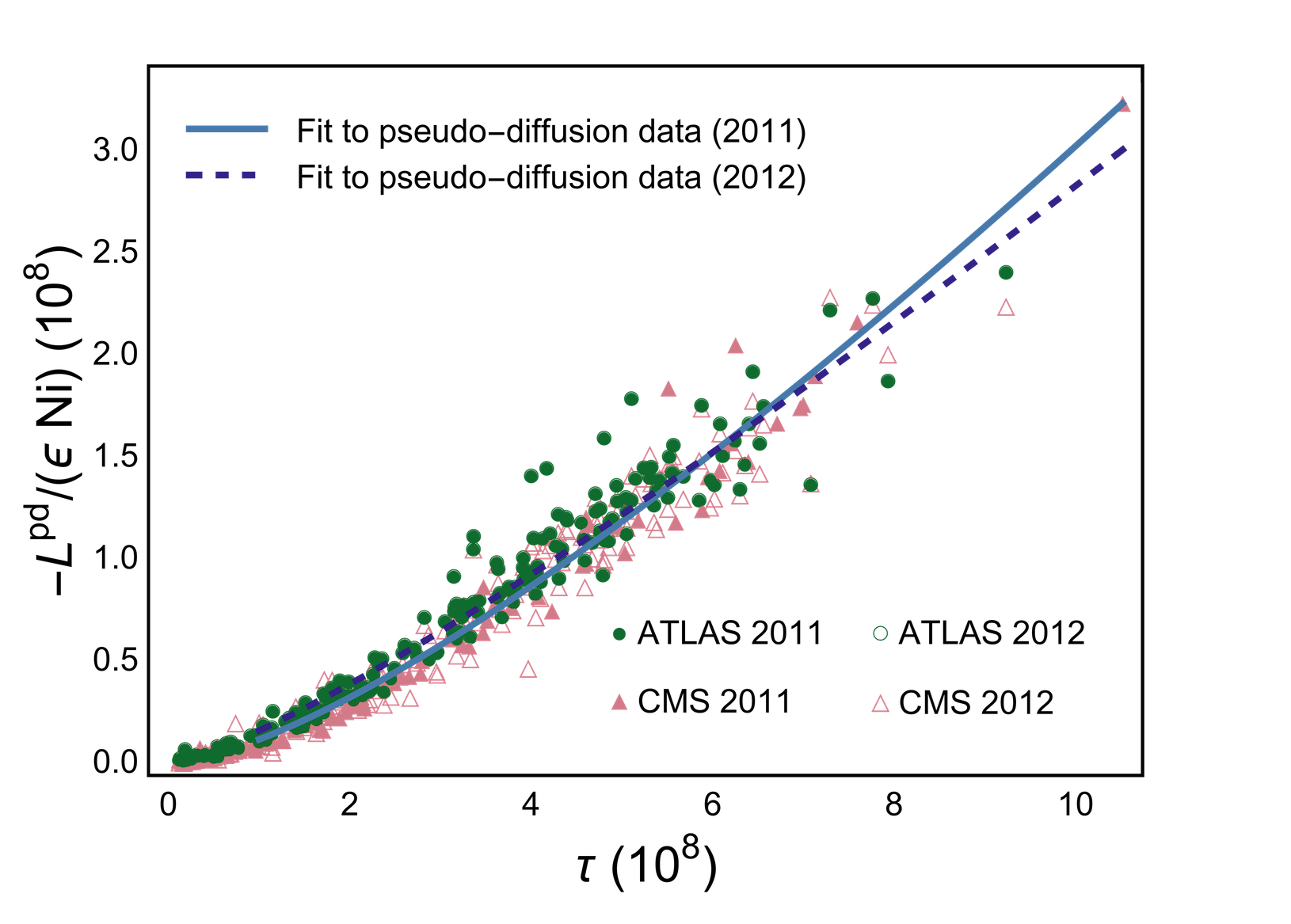} \\
      \includegraphics[width=0.49\linewidth,clip=]{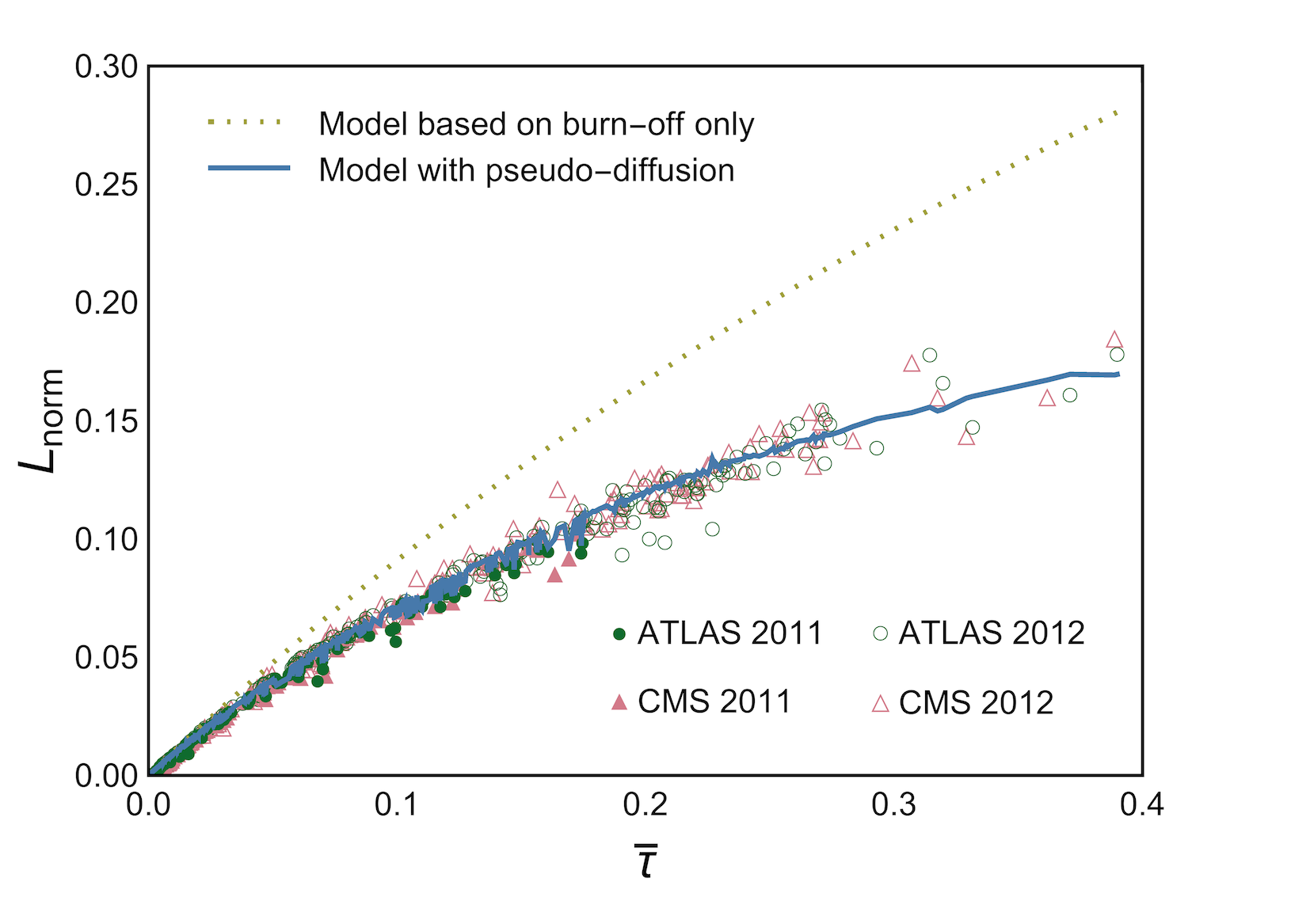} &
      \includegraphics[width=0.49\linewidth,clip=]{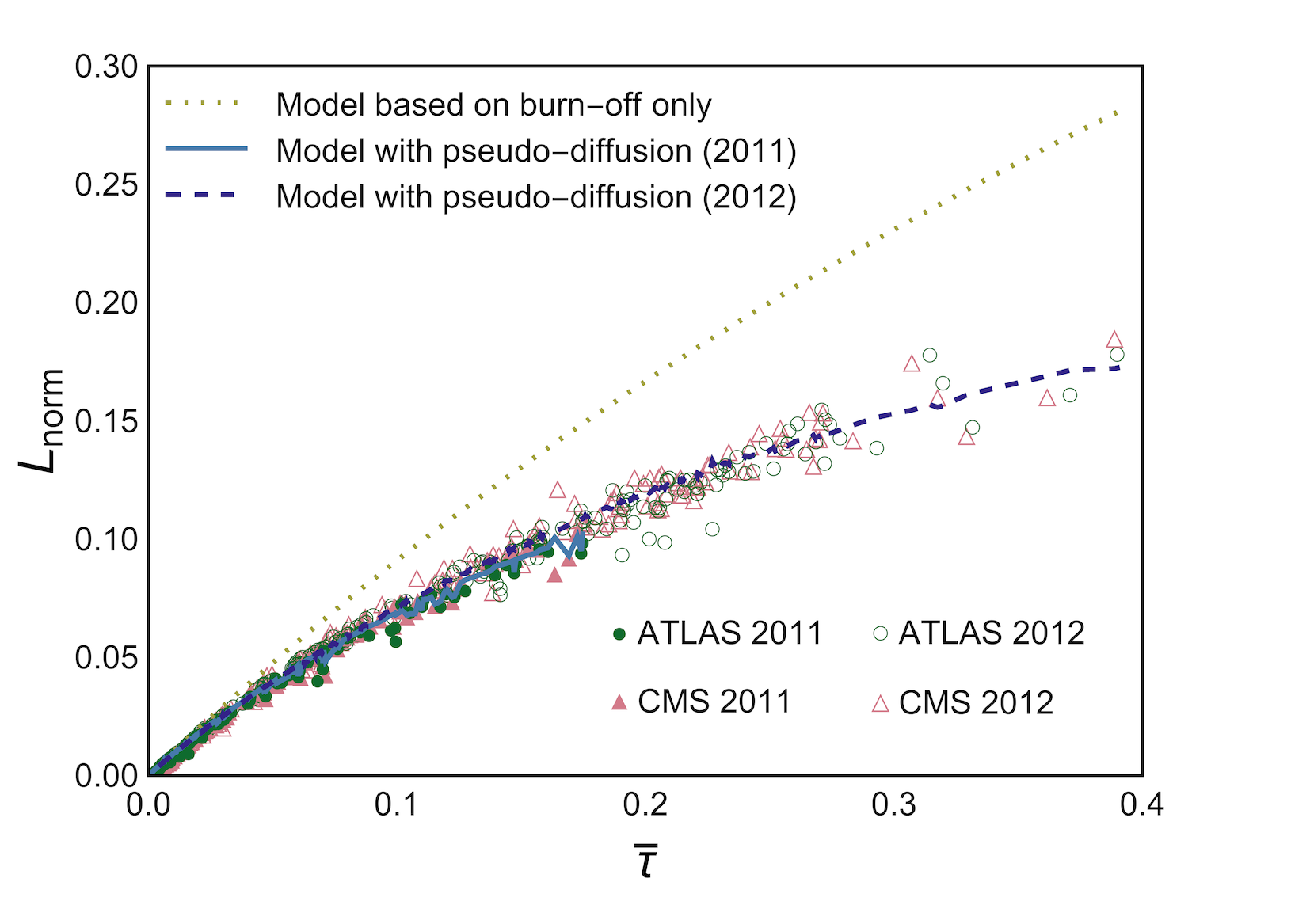} \\
    \end{tabular}
  \end{center}
  \caption{Pseudo-diffusive contribution to the integrated luminosity as a function of the turn number (upper row). Normalised integrated luminosity as a function of the normalised time shown together with models based on burn off and pseudo-diffusive effect (lower row). The model has been fitted using the complete set of 2011 and 2012 data (left), and separately year by year (right). Note that for 2011 we have $\bar{\tau}<0.2$. The agreement between the proposed model with pseudo-diffusive effects and the experimental data is remarkable.} 
  \label{modelfit}
\end{figure}

As a first investigation, the pseudo-diffusion model has been fitted to the complete Run~1 dataset. This is shown on the left side of Fig.~\ref{modelfit}, and the values of the fit parameters including the associated errors are reported in Table~\ref{fit_par_long}. 

From Fig.~\ref{modelfit} it is clear that including the pseudo-diffusive effects is very efficient to recover a nice agreement between model predictions and measured data, as the discrepancy between the burn off-only model and the measurements is rather large if $\bar{\tau} > 0.1$. Furthermore, it is important to note that the fit parameters of the pseudo-diffusive component correspond, given that all parameters are positive, to a situation in which a stable region exists in phase space for an arbitrarily long time. Note also that $\kappa$ is extremely close to $2$, the estimate given in the proof of the Nekhoroshev theorem. 

The pseudo-diffusive effect on a yearly basis is shown in the right plot of Fig.~\ref{modelfit}, and a difference between the two years is seen, which does not exceed $20~\%$. Careful inspection reveals that the same difference exists in the data, thus confirming that the model reproduces closely the features of the dataset. Note that $L_{\rm norm}$ is supposed to be independent of the beam parameters, hence it should be the same for 2011 and 2012, but this only holds for the burn off part. Indeed, in Eq.~\eqref{Lpdfitmodel} the pseudo-diffusive part is scaled by a factor $-\varepsilon N_{\rm i}$ which depends on the beam parameters, while the DA model parameters in the integrand are sensitive to the specific ring conditions. The parameter values for the yearly fits are also given in Table~\ref{fit_par_long}. Note that now $D_\infty < 0$ for 2011, exactly like in the non-integrated case, again implying that the phase space is dominated by a situation of global chaoticity. This difference in behaviour between 2011 and 2012 runs is still lacking an explanation based on considerations linked with the run conditions. 

\begin{table}[htb]
\centering
\caption{Summary of the fit parameters and associated errors for $L^{\rm pd}(\tau)$, for different data subsets. The error on the fit parameters is estimated using the BCa interval. For the parameters shown in italic, the fit estimate lies outside the Bootstrap 90~\% confidence interval. In that case the standard deviation of the bootstrap realisations is used instead (see \ref{app:difficulties}).}
\begin{tabular}{lccccc}
\hline
& $D_\infty$ & $b$ & $\kappa$ & $R^2_\text{adj} [\%] $ \\ \hline
Run~1 (2011+2012)
	& $\mathit{0.44 \pm 0.54}$	& $\mathit{460 \pm 110}$	& $\mathit{1.92 \pm 0.31}$	& 96.433\\
$\kappa = 2$ \qquad
	& $0.497_{-0.054}^{+0.095}$	& $556_{-37}^{+ 20}$	& --	& 96.440\\
$D_\infty = 0$ \qquad
	& --	& $177_{-43}^{+30}$	& $1.517_{- 0.094}^{+ 0.052}$	& 96.434\\
$\,\kappa = 2$, $D_\infty = 0$ \qquad
	& --	& $\mathit{740.0 \pm 1.1}$	& --	& 96.208\\ \hline
2011 \qquad
	& $-0.43_{- 0.14}^{+ 0.38}$	& $350_{- 80}^{+ 150}$	& $1.68_{- 0.13}^{+ 0.16} $
	& 97.835\\
$\kappa = 2$ \qquad
	& $-0.03_{- 0.13}^{+ 0.10} $	& $757_{- 35}^{+ 49}$	& --	& 97.847\\
$D_\infty = 0$ \qquad
	& --	& $830_{- 200}^{+ 370}$	& $2.04_{- 0.09}^{+ 0.13}$	& 97.848\\
$\,\kappa = 2$, $D_\infty = 0$ \qquad
	& --	& $744.0_{- 1.8}^{+ 1.6}$	& --	& 97.857\\ \hline
2012 \qquad
	& $\mathit{0.82 \pm 0.52}$	& $\mathit{560 \pm 114}$	& $\mathit{2.08 \pm 0.35}$	& 95.746\\
$\kappa = 2$ \qquad
	& $0.77_{- 0.06}^{+ 0.13}$	& $455_{- 49}^{+ 21}$	& --	& 95.754\\
$D_\infty = 0$ \qquad
	& --	& $81_{- 26}^{+ 15}$	& $1.25_{- 0.13}^{+ 0.06}$	& 95.737\\
$\,\kappa = 2$, $D_\infty = 0$ \qquad
	& --	& $\mathit{738.2 \pm 1.4}$	& --	& 95.166 \\
\hline
\label{fit_par_long}
\end{tabular}
\end{table}

\begin{figure}[htb]
  \begin{center}
      \begin{tabular}{@{}c@{}@{}c@{}}
        \includegraphics[width=0.49\linewidth,clip=]{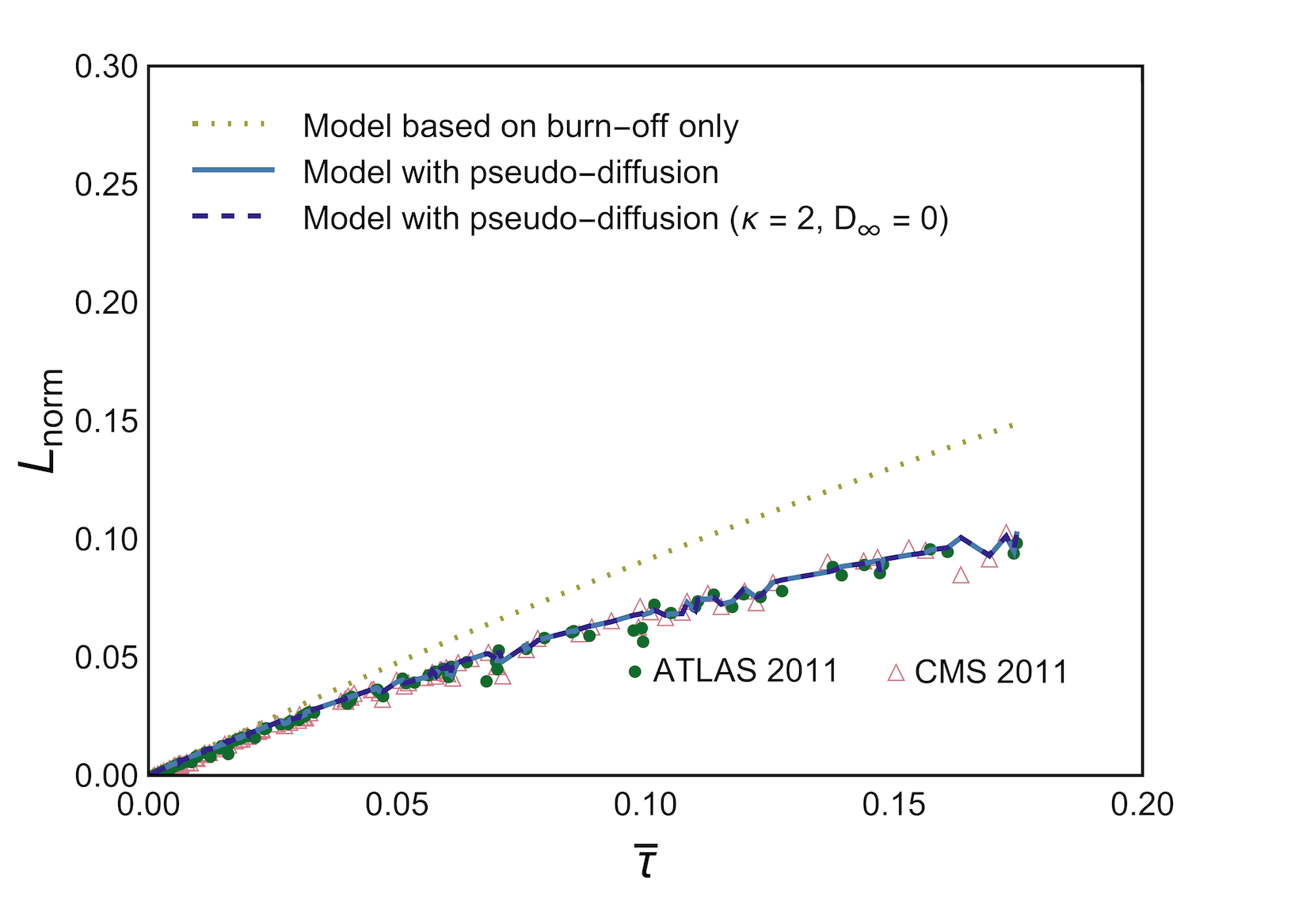} &
        \includegraphics[width=0.49\linewidth,clip=]{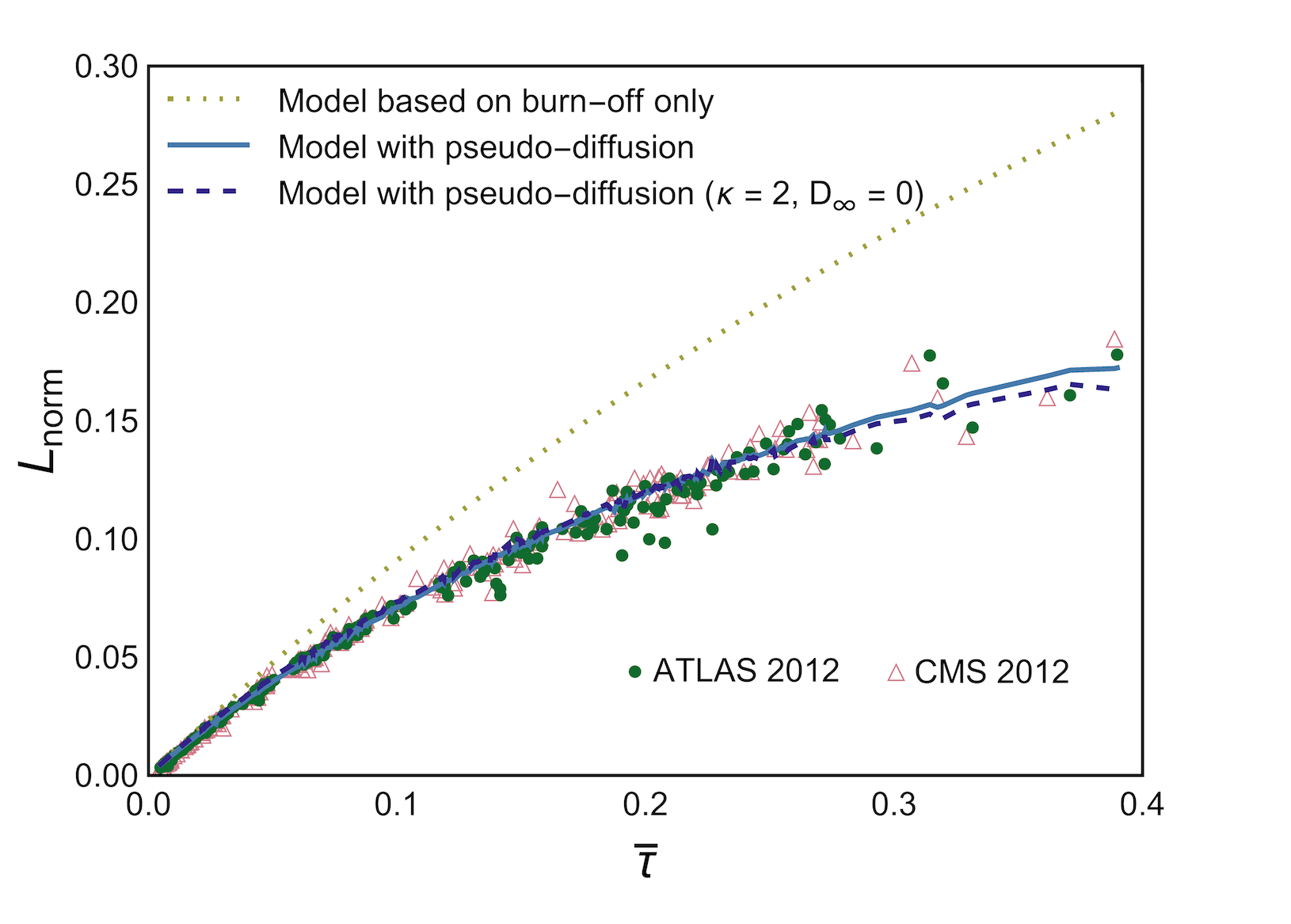} \\
    \end{tabular}
  \end{center}
  \caption{Different models of the pseudo-diffusive contribution to the luminosity for proton beam data from 2011 (left) and 2012 (right). The curves for the two-parameter fits are not shown, as they are almost perfectly superimposed with the curve for three-parameter fit.} 
  \label{modelfitnew}
\end{figure}

Based on the considerations made in \ref{app:difficulties} and on the observations made when fitting the non-integrated luminosity, it is possible to carry out a refined analysis of the Run~1 data by reducing the number of fit parameters. The resulting parameters and the corresponding errors are listed in Table~\ref{fit_par_long}, while the corresponding plots for 2011 and 2012 are shown in Fig.~\ref{modelfitnew}. 

The ranking of the various fit types for a given dataset can be performed based on the corresponding value of $R^2_\text{adj}$. This allows to state that, e.g. for the combined dataset (2011+2012) or 2012 alone, the model in which both $D_{\infty}$ and $\kappa$ are fixed provides the worst performance. For the 2011 dataset, however, all four models behave very similarly. All in all, this suggests that the best strategy is to compare the fit with three parameters to a variant with two parameters only and $\kappa$ determined based on the theoretical estimate. These considerations are confirmed by Fig.~\ref{modelfitnew}, where no difference is visible between the four types of fit for the 2011 data, while for 2012 only the model where both $\kappa=2$ and $D_\infty=0$ is different from the others, with a difference not exceeding $5~\%$, but also performs worse according to the value of $R^2_\text{adj}$.

A comparison of the fit parameters for the corresponding cases reported in Tables~\ref{fit_par_unint} and~\ref{fit_par_long} shows that the values are compatible, within the errors, for the case of three-parameter fit, while the compatibility degrades as the number of fit parameters is reduced, the case with fixed $\kappa$ being more compatible between the non-integrated and integrated luminosity models, than that with $D_{\infty}$. This confirms once more that fixing $\kappa$ is the best option among those with reduced fit parameters.
\section{Optimal physics fill duration}\label{sec:fill_duration_data}
As discussed in Ref.~\cite{lumi_Part_I}, the optimisation of the performance of a circular collider can be performed by maximising the yearly integrated luminosity given by 
\begin{equation}
L_{\rm tot, norm}^{\rm bo}(\tau) = \frac{\mathcal{T}}{\tau_{\rm ta}+\tau} 
\, L_{\rm norm}^{\rm bo} (\tau) \, ,
\label{lumitotnorm}
\end{equation}
where $\tau_{\rm ta}$ is the turnaround time, i.e. the time between the end of a physics fill and the beginning of the next one, $\tau$ is the fill length that should be optimised, and $\mathcal{T}$ is the total time for physics over one year and only the burn off has been taken into account is considered. The optimal fill length $\tau_{\rm fill}$ can be obtained by setting to zero the derivative of $L_{\rm tot, norm}^{\rm bo}$. 

\begin{figure}[htb]
  \begin{center}
      \begin{tabular}{@{}c@{}@{}c@{}}
      \includegraphics[width=0.49\linewidth,clip=]{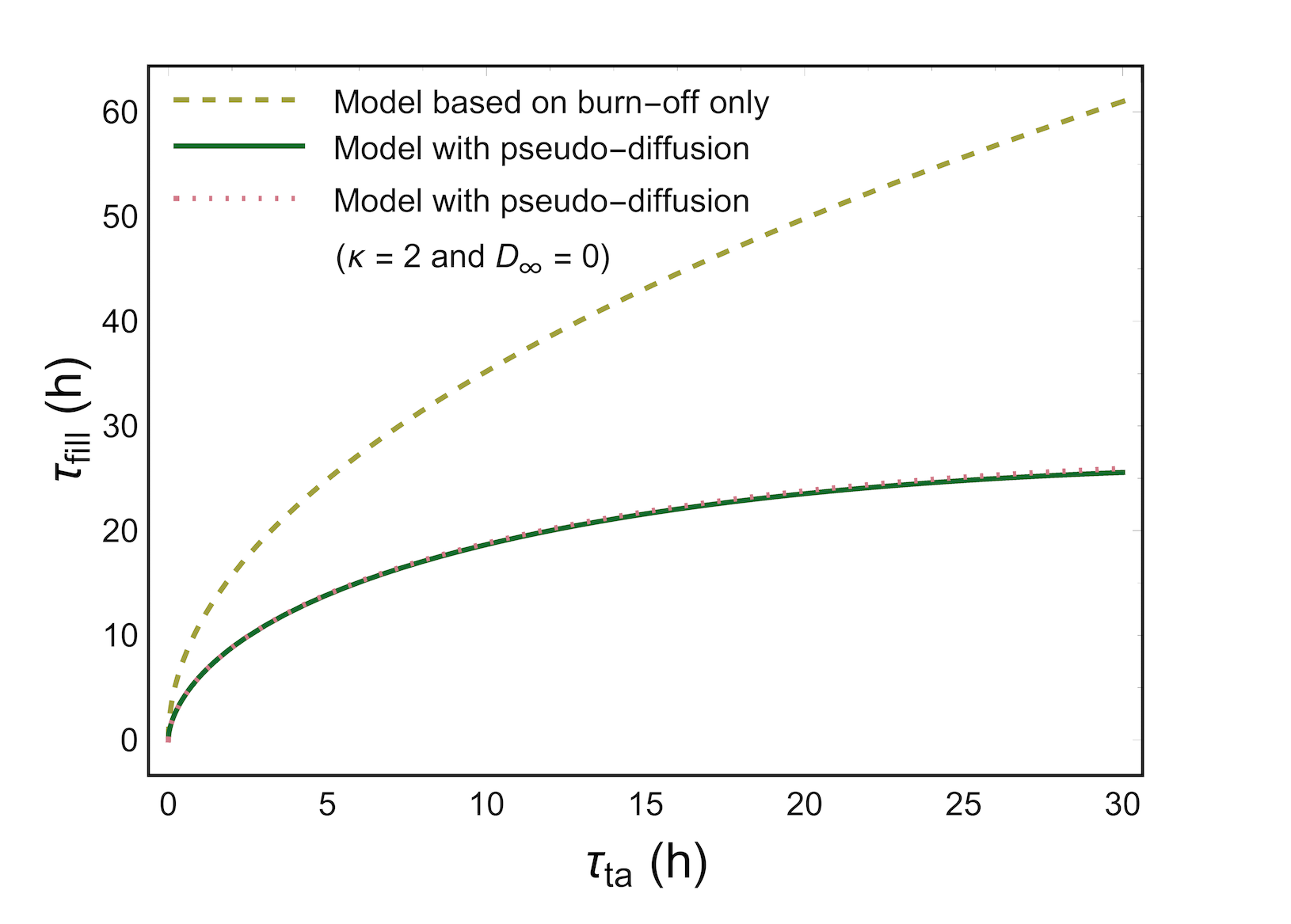} &
      \includegraphics[width=0.49\linewidth,clip=]{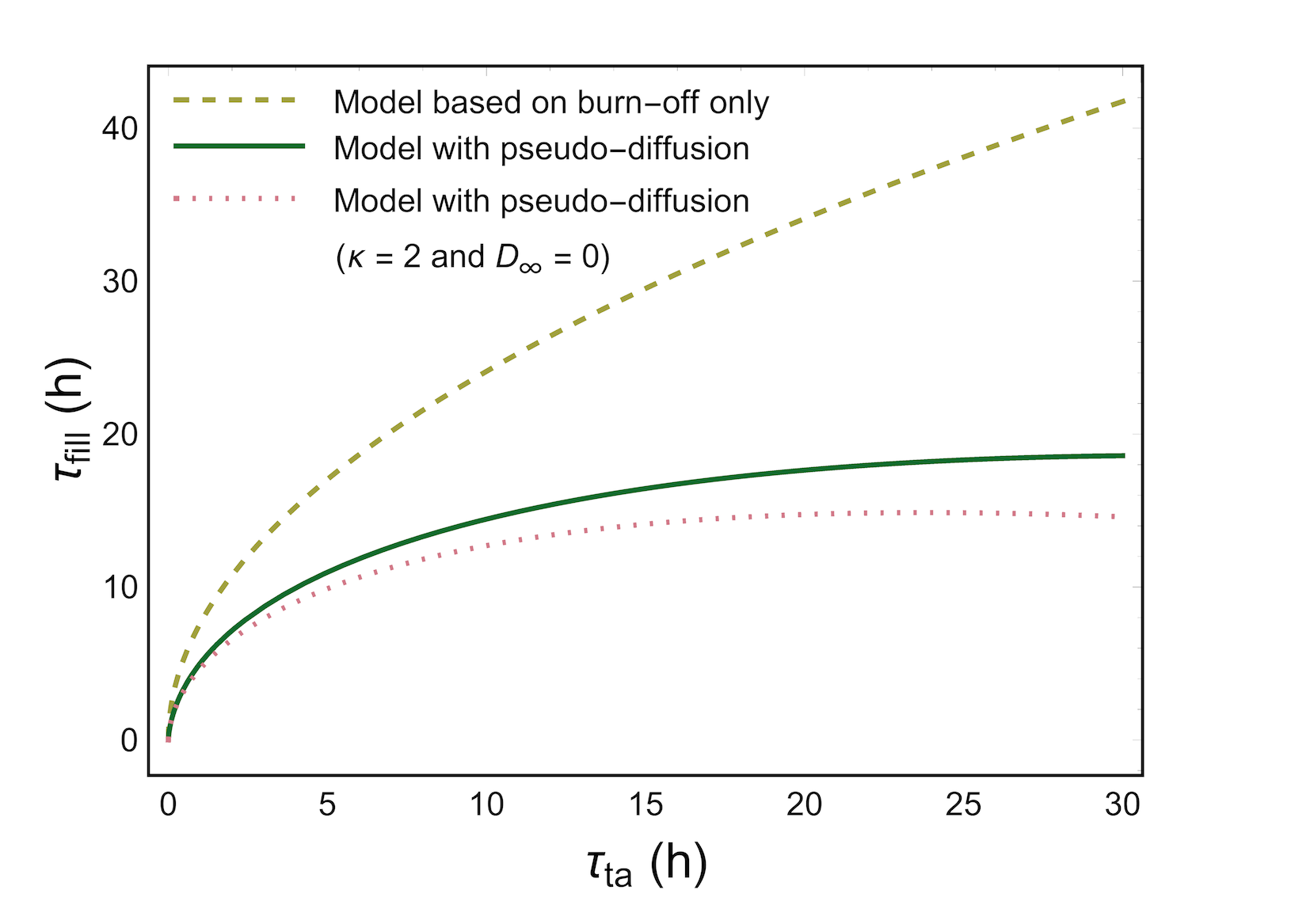} \\
    \end{tabular}
  \end{center}
  \caption{Optimal fill length $\tau_\text{fill}$ as a function of the turnaround time $\tau_\text{ta}$ for 2011 (left) and 2012 (right). The curves for the two-parameter fits are not shown, as they are almost perfectly superimposed with the curve for three-parameter fit. The parameters of the models used to estimate $\tau_{\rm fill}$ are those from Table~\ref{fit_par_long}.} 
  \label{optimalfill}
\end{figure}

Of course, the term $L^{\rm pd}$ changes the conclusions concerning the optimal fill time and an approximate expression reads~\cite{lumi_Part_I}
\begin{equation}
\tau_{\rm fill} \approx
  \tau_{\rm fill}^{\rm bo} -
  \frac{
    \dot{L}_{\rm tot, norm}^{\rm pd}(\tau_{\rm fill}^{\rm bo})
  }%
  {
    \ddot{L}_{\rm tot, norm}^{\rm bo}(\tau_{\rm fill}^{\rm bo})
    +\ddot{L}_{\rm tot, norm}^{\rm pd}(\tau_{\rm fill}^{\rm bo})
  } 
\label{taufill}
\end{equation}

and it is possible to use Eq.~\eqref{taufill} to estimate the optimal physics fill duration for our models, as shown in Fig.~\ref{optimalfill}. Note that there is a clear difference between the case where only burn off is taken into account, and that where pseudo-diffusive effects are included. The relation between the various models used to derive $\tau_{\rm fill}$ is similar to that for the fit of $L^\text{pd}\,$, namely, for 2011 all four cases behave the same, while for 2012 only the case where two parameters are fixed ($\kappa=2$ and $D_\infty=0$) is different from the others.

\begin{figure}[htb]
  \begin{center}
      \includegraphics[width=0.59\linewidth,clip=]{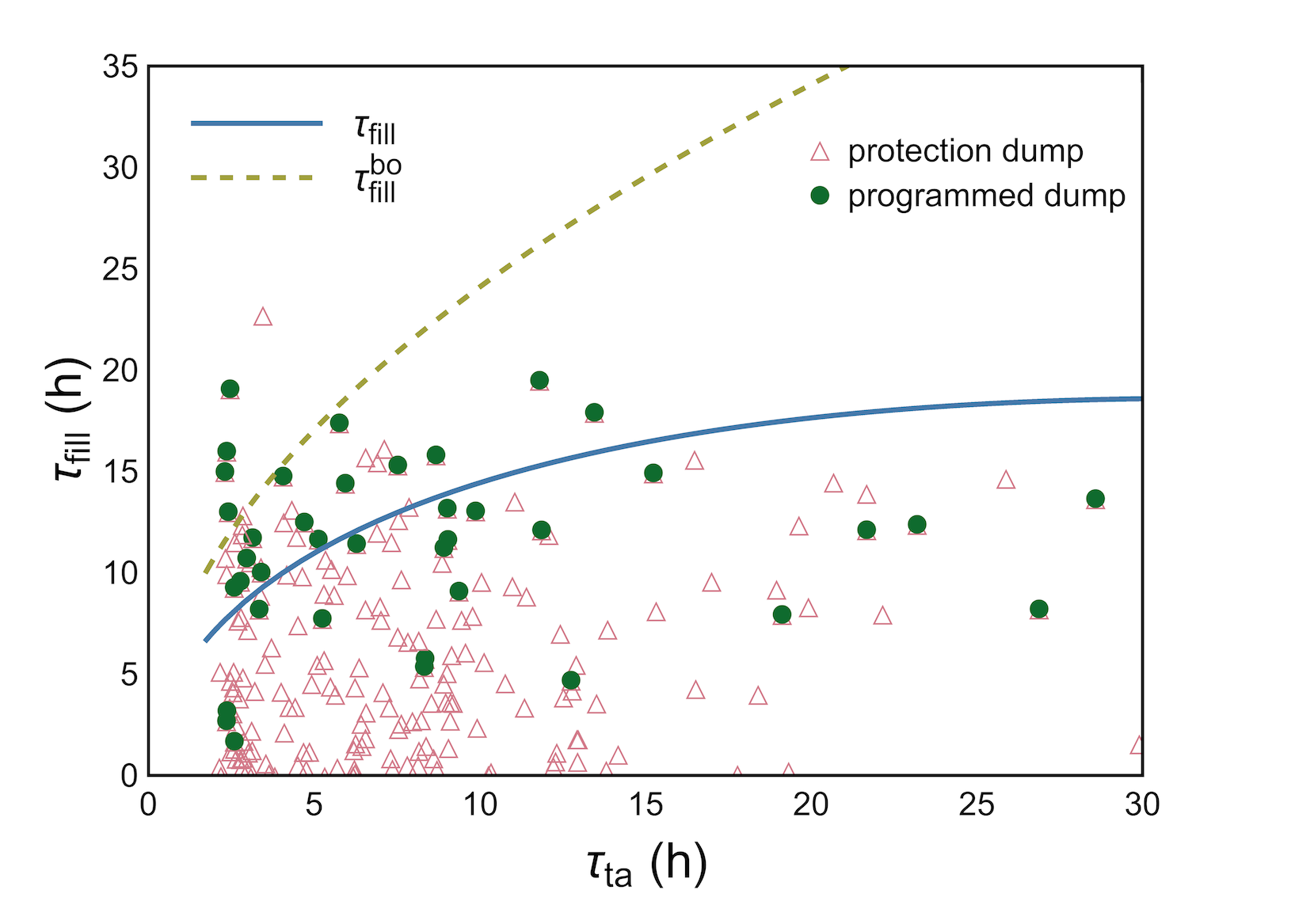} 
  \end{center}
\caption{Length of physics fill as a function of the turnaround time for all stable physics fills from 2012. The fills are compared to the estimate for the optimal fill length with and without pseudo-diffusion.}
  \label{filldata}
\end{figure}

Then, we carried out the comparison of the estimate for the optimal fill length with the actual fill duration during Run~1, focusing on the situation for the year 2012. The distribution of the actual values of $\tau_{\rm fill}$ as a function of  $\tau_{\rm ta}$ is shown in Fig.~\ref{filldata}. The data considered includes all fills for high-luminosity physics, excluding all other cases, e.g. special runs, commissioning periods. Note, also, that $\tau_{\rm ta}$ for a given physics fill is computed as the time between the end of the previous physics fill and the beginning of that under consideration. All these data have been divided into two groups: a class in which the end of the fill for physics is controlled by operation, so-called {\em programmed dump}, and a class in which the end of the fill for physics is triggered by the machine protection system, so-called {\em protection dump}. The relevance of this classification is that the first class allows for performing an optimisation of the fill length, while the latter does not. It is also worth mentioning that even in the case of the first class, there are some situations in which a beam dump is indeed triggered by operation, but in view of preventing a protection dump. It is, e.g. the case when the cryogenic conditions are going to be lost in a short while and the operator dumps the beams before a genuine protection dump occurs. This special subset of the first class explains the cases of programmed dumps with rather short $\tau_{\rm fill}$. 

Figure~\ref{filldata} shows a number of interesting features: $\tau_{\rm ta}$ is always larger than $\sim 2$~h, the minimum turnaround time based on the performance of the LHC hardware; $\tau_{\rm ta}$ can reach rather high values, which indicate a fault that occurred in between fills for physics; for the class of programmed dump the length of the physics fill is clearly different from the corresponding optimal fill length, either because the physics fill is too short or because it is too long; for the class of protection dump some fill are also too long with respect to the optimal duration. While the proposed approach is very useful to qualify whether a single fill is optimal with respect to its duration, the overall optimisation of the yearly performance, in terms of integrated luminosity, of a collider is a much more complex task.  
\section{Conclusions\label{sec:conclusions}}
The models proposed in the companion paper~\cite{lumi_Part_I} have been benchmarked against the data from the LHC Run~1, with special emphasis on the years 2011 and 2012. The inaccuracy to reproduce the LHC data using only burn off has been confirmed by the analysis made, while the proposed models showed a remarkable power in reproducing and describing the observed behaviours of luminosity as a function of time and of integrated luminosity. 

A detailed discussion of the potential numerical issues related with the proposed fitting models is presented in \ref{app:difficulties}. It is also shown that the difficulties can be efficiently resolved by reducing the number of fit parameters, by using the estimates provided by the proof of Nekhoroshev theorem for some of them. 

Given the encouraging results of the analyses reported in this paper, the data from Run~2 will be considered next. In fact, the higher beam energy that characterises the proton physics in Run~2 opens a new domain in terms of beam behaviour, such as strong longitudinal emittance damping due to synchrotron radiation as well as a burn-off dominated regime.

Finally, it is worth stressing that while in this paper the approach has been to fit the model parameters to measured data, in future the analysis can be shifted to using numerical simulations to provide the input about the dynamic aperture evolution, which is needed in the proposed models, to verify the agreement with observations from the LHC. In this way the descriptiveness of the proposed approach might turn into predictiveness, which could be used to assess the performance for future colliders and in particular the luminosity upgrade of the LHC.
\section*{Acknowledgements}
One of the authors (MG) would like to thank A.~Bazzani for interesting discussions as well as G.~Arduini for interesting and stimulating remarks and comments. We would like to thank R.~De~Maria and G.~Iadarola for help with the data extraction.
\section*{References}
\clearpage
\appendix
\section{Comments on numerical aspects of the proposed fit model}\label{app:difficulties}
The proposed model for luminosity evolution implies a fit of the experimental data to derive the values of the model parameters. Special care is needed for performing the fit, as, in general, internal dependencies between the model parameters have been observed and need to be considered in detail. Moreover, for the specific case of the model for the integrated luminosity, the integral function that is fitted to the data might be too sensitive to small changes in the model parameters, thus adding more challenges. 

\begin{figure}[htb]
  \begin{center}
    \begin{tabular}{@{}c@{}@{}c@{}}
      \includegraphics[width=0.49\linewidth,clip=]{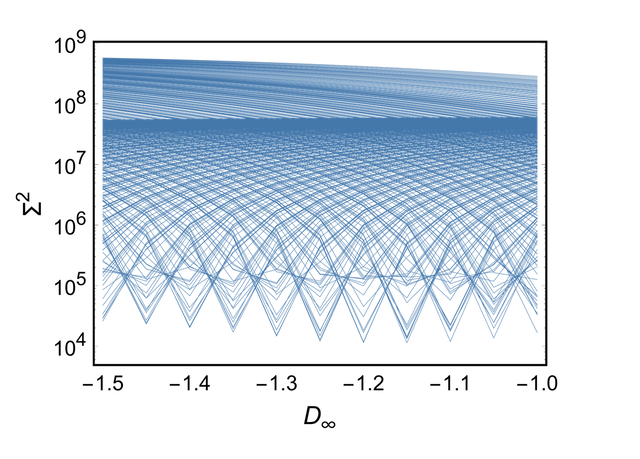} &
      \includegraphics[width=0.49\linewidth,clip=]{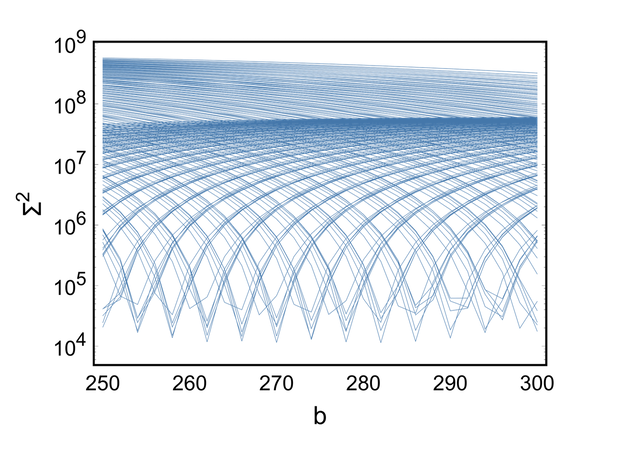} \\
    \end{tabular}
    \includegraphics[width=0.49\linewidth,clip=]{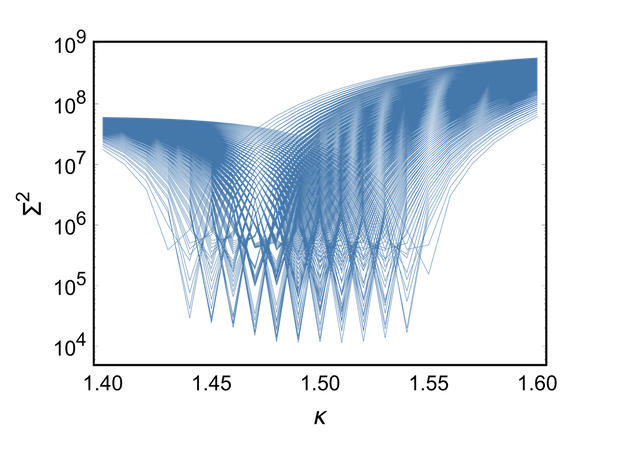}
  \end{center}
  \vspace*{-\baselineskip}
  \caption{Behaviour of $\Sigma^2$ for the fit of the model applied to fill 2240 as a function of the fit parameters. In each plot every curve represents the influence of the listed parameter on $\Sigma^2$, while the other two parameters are kept constant. Different curves represent different values of the other two parameters. The minima of $\Sigma^2$ are approximately degenerate as several optimal points (in parameter space) exist with similar values of $\Sigma^2$.} 
  \label{residues}
\end{figure}

To illustrate this phenomenon and to analyse better the behaviour of the fit, the sum of the residues squared, which is the figure of merit used by the fitting algorithm
\begin{equation}\label{eq:RSS}
\Sigma^2(D_{\infty}, b , \kappa) =
	\sum_i \left [ y_i - f(\tau_i,D_{\infty}, b , \kappa) \right ]^2 \, ,
\end{equation}
has been considered, $y_i$ being the measured data and $f$ the fit model. An internal dependency between model parameters would manifest in terms of a degenerate minimum, i.e. changing one parameter while adapting another one would not change significantly $\Sigma^2$. The fill 2240 of the year 2011 has been used as a test case as it is particularly well behaved. The fitting algorithm provided the following values of the parameters, $D_\infty=-1.22$, $b=286.79$, and $\kappa=1.51$, and, whenever a scan around these values is performed, the behaviour of $\Sigma^2$ looks like in the plots of Fig.~\ref{residues}. An approximate degeneration is visible, showing that several combinations of fit parameters can give similarly low values of $\Sigma^2$.

This essential observation leads to the conclusion that decreasing the number of fit parameters might be advisable. A natural option is to fix $\kappa$ to the value provided by the estimate in the proof of Nekhoroshev theorem~\cite{nekhor}, i.e. $\kappa=(d+1)/2$, where $d$ is the number of degrees of freedom of the system under consideration. Other possibilities will be considered and discussed in the next sections.

Another important aspect to consider is the estimate of the error associated with the fit parameters. Such an estimate is closely linked with that of the error associated to the luminosity measurement and crucially depends upon it. Given that it is not so easy to provide a reliable estimate for the total error on luminosity, the so-called Bootstrap Method~\cite{bootstrap,bootstrap2} has been applied. Here, the population of residues of the fit is used as a distribution to generate 10\,000 realisations with replacements, that are again fitted to the model. The set of realisations of each parameter will then be distributed around the original results of the fit procedure. Therefore, the parameter's distribution can provide an estimate of the error associated with the model parameter by means of the bias-corrected accelerated (BCa) Bootstrap confidence interval~\cite{bootstrap2} at $1\sigma$. For the studies that exhibit degeneracy in the parameter space, it is often the case that the fitted parameter value lies outside the BCa interval, even at 90~\%. In such case the BCa interval calculation is less meaningful and we estimate the error by the standard deviation of the distribution of the fit parameters over the realisations instead.

\begin{thebibliography}{99}
%
\bibitem{lumi_Part_I} M.~Giovannozzi, F.F.~Van~der~Veken, ``Description of the luminosity evolution for the CERN LHC including dynamic aperture effects. Part I: the model'', submitted to publication to Nucl. Instr. Meth. Phys. Res. A.
%
\bibitem{dynap1} M.~Giovannozzi, W.~Scandale, E.~Todesco, ``Prediction of long-term stability in large hadron colliders'', Part. Accel. {\bf 56} 195, 1996.
%
\bibitem{dynap2} M.~Giovannozzi, W.~Scandale, E.~Todesco, ``Dynamic aperture extrapolation in presence of tune modulation'', Phys. Rev. E {\bf 57} 3432, 1998.
%
\bibitem{loginvb-b} M.~Giovannozzi, E.~Laface, ``Investigations of Scaling Laws of Dynamic Aperture with Time for Numerical Simulations including Weak-Strong Beam-Beam Effects'', TUPPC086, IPAC12 proceedings, p.~1359, 2012.
%
\bibitem{lossesPRSTAB} M.~Giovannozzi, ``Proposed scaling law for intensity evolution in hadron storage rings based on dynamic aperture variation with time'', Phys. Rev. ST Accel. Beams {\bf 15} 024001, 2012.
%
\bibitem{DAexp_nekor} E.~Maclean, M.~Giovannozzi and R.~Appleby, ``A novel method to measure the extent of the stable phase-space region of proton synchrotrons using Nekhoroshev-like scaling laws'', submitted for publication.
%
\bibitem{Lumi_fit} M.~Giovannozzi, C.~Yu, ``Proposal of an Inverse Logarithm Scaling Law for the Luminosity Evolution'', TUPPC078, IPAC12 proceedings, p.~1353, 2012.
%
\bibitem{IPAC14} M.~Giovannozzi, ``Simple models describing the time-evolution of luminosity in hadron colliders'', TUPRO009, IPAC14 proceedings, p.~1017, 2014.
%
\bibitem{MC1} M.Crouch,``Luminosity performance limitations due to the beam-beam interaction in the LHC'', Thesis, University of Manchester, U.K.
%
\bibitem{MC2} M.~Crouch, T.~Pieloni, R.B.~Appleby, J.~Barranco-Garc\'{i}a, X.~Buffat, M.~Giovannozzi, E.H.~Maclean, B.D.~Muratori, C.~Tambasco,``Dynamic aperture studies of long-range beam-beam interactions at the LHC'', THPAB056, IPAC17 proceedings, p.~3840, 2017.
%
\bibitem{Herr} W. Herr, ``Concept of Luminosity'', CAS - CERN Accelerator School: Intermediate Course on Accelerator Physics, Zeuthen, Germany, 15 - 26 Sep 2003, p.361 (CERN-2006-002).
%
\bibitem{lpc} https://lpc.web.cern.ch/Default.htm
%
\bibitem{inel1} The ATLAS and CMS Collaborations, ``Expected pile-up values at HL-LHC'', ATL-UPGRADE-PUB-2013-014, 2013.
%
\bibitem{inel2} D.~Contardo, private communication, 3 December 2014.
%
\bibitem{RunI_1} M.~Lamont, ``The LHC from Commissioning to Operation'', MOYAA01, IPAC11 proceedings, p.~11, 2011.
%
\bibitem{RunI_2} S.~Myers, ``The First Two Years of LHC Operation'', MOXBP01, IPAC12 proceedings, p.~1, 2012.
%
\bibitem{RunI_3} J.~Wenninger, R.~Alemany-Fernandez, G.~Arduini, R.W.~Assmann, B.J.~Holzer, E.B.~Holzer, V.~Kain, M.~Lamont, A.~Macpherson, G.~Papotti, M.~Pojer, L.~Ponce, S.~Redaelli, M.~Solfaroli Camillocci, J.A.~Uythoven, W.~Venturini Delsolaro, ``Operation of the LHC at High Luminosity and High Stored Energy'', THPPP018,  IPAC12 proceedings, p.~3767, 2012.
%
\bibitem{RunI_4} M.~Lamont, ``The First Years of LHC Operation for Luminosity Production'', MOYAB101, IPAC13 proceedings, p.~6, 2013.
%
\bibitem{MG_note} M.~Giovannozzi, ``Some considerations on the p-p performance of the LHC during Run I'', CERN-ACC-NOTE-2013-0039. 
%
\bibitem{data_storage} {https://lhc-statistics.web.cern.ch/LHC-Statistics/index.php}
%
\bibitem{spsblowup} G.~Arduini, ``Performance reach in the LHC for 2012'', in Proceedings of Chamonix 2012 workshop on LHC Performance, CERN-2012-006, p.~189, 2012.
%
\bibitem{lumi_Part_III} M.~Giovannozzi, F.F.~Van~der~Veken, ``Description of the luminosity evolution for the CERN LHC including dynamic aperture effects. Part III: application to Run~2 data'', in preparation.
%
\bibitem{Inst2012} T.~Pieloni, G.~Arduini, R.~Giachino, W.~Herr, M.~Lamont, E.~M\'etral, N.~Mounet, G.~Papotti, B.~Salvant, J.~Wenninger, X.~Buffat, S.~M.~White, ``Observations of two-beam instabilities during the 2012 LHC physics run'', TUPFI034, IPAC13 proceedings, p.~1418, 2013.
%
\bibitem{OMC1} R.~Tom\'as, T.~Bach, R.~Calaga, A.~Langner, Y.~I.~Levinsen, E.~H. Maclean, T.~H.~B.~Persson, P.~K.~Skowronski, M.~Strzelczyk, G.~Vanbavinckhove, and R.~Miyamoto, ``Record low $\beta$-beating in the LHC'',  Phys. Rev. ST Accel. Beams {\bf 15}, 091001 (2012). 
%
\bibitem{OMC2} T.~Persson, F.~Carlier, J.~Coello de Portugal, A.~Garcia-Tabares Valdivieso, A.~Langner, E.~H.~Maclean, L.~Malina, P.~Skowronski, B.~Salvant, and R.~Tom\'as, ``LHC optics commissioning: A journey towards $1~\%$ optics control'',  Phys. Rev. Accel. Beams {\bf 20}, 061002 (2017).
%
\bibitem{Jorg} J.~Wenninger, ``Simple models for the integrated luminosity'', CERN-ATS-Note-2013-033 PERF, 2013.
%
\bibitem{bootstrap} B.~Efron, ``Bootstrap Methods: Another Look at the Jackknife'', Ann. Stat. {\bf 7}, 1, 1979.
%
\bibitem{bootstrap2} J.~Fox, ``Applied Regression Analysis and Generalized Linear Models'', SAGE Publications (Mar., 2015).
%
\bibitem{nekhor} N.~Nekhoroshev, ``An exponential estimate of the time of stability of nearly-integrable Hamiltonian systems'', Russ. Math. Surv. {\bf 32} 1, 1977.
%
\end{thebibliography}
\end{document}